\documentclass[showpacs,showkeys,preprintnumbers,prd,amsmath,amssymb,floats,floatfix]{revtex4}
\usepackage{amsmath,amssymb,bm,natbib}
\usepackage[paperwidth=210mm, paperheight=297mm,centering,hmargin=3cm,vmargin=3.5cm]{geometry}
\usepackage{epstopdf}
\usepackage{hyperref}
\usepackage{color}
\usepackage{booktabs}
\usepackage{graphicx}
\usepackage{multirow}
\usepackage{slashed}
\usepackage{comment}
\usepackage{float}
\makeatletter

\newcommand{\Rmnum}[1]{\expandafter\@slowromancap\romannumeral #1@}

\makeatother
\begin{document}
\def\bib{\bibitem}
\def\be{\begin{equation}}
\def\ee{\end{equation}}
\def\beq{\begin{equation}}
\def\eeq{\end{equation}}
\def\beqar{\begin{eqnarray}}
\def\eeqar{\end{eqnarray}}
\def\barr{\begin{array}}
\def\earr{\end{array}}
\def\dis{\displaystyle}
\def\lsim{\:\raisebox{-0.5ex}{$\stackrel{\textstyle<}{\sim}$}\:}
\def\gsim{\:\raisebox{-0.5ex}{$\stackrel{\textstyle>}{\sim}$}\:}
\def\tilh{\tilde{h}}
\def\and{\qquad {\rm and } \qquad}
\def\vev{\small \em {\it v.e.v. }}
\def\p{\partial}
\def\ga{\gamma^\mu}
\def\slp{p \hspace{-1ex}/}
\def\sleps{ \epsilon \hspace{-1ex}/}
\def\slk{k \hspace{-1ex}/}
\def\slq{q \hspace{-1ex}/\:}
\def\prd#1{Phys. Rev. {\bf D#1}}
\def\etal{ {\it et al.} }
\def\ie{ {\it i.e.} }
\def\eg{ {\it e.g.} }

\title{Constraining capability of $Z \gamma h$ production at the ILC}
 
\author{\bf Sher Alam\textsuperscript{a}\footnote{iasher0e@gmail.com}, \bf Subhasish Behera\textsuperscript{b}\footnote{subhasish@iitg.ernet.in}, \bf Satendra Kumar\textsuperscript{c}\footnote{satendrak@iitgn.ac.in}, \bf Shibananda Sahoo\textsuperscript{b}\footnote{shibananda@iitg.ernet.in}}

\affiliation{\textsuperscript{a}
Center of High Energy Physics, University of Punjab, Pakistan,
Japan-Pakistan Collaboration, 448-18 Ozone, Japan}

\affiliation{\textsuperscript{b}
Department of Physics,
Indian Institute of Technology Guwahati,
Guwahati 781 039, India}

\affiliation{\textsuperscript{c}
Indian Institute of Technology Gandhinagar,
Gandhinagar 382 355, India}

\begin{abstract} 
Higgs boson couplings with gauge bosons are probed through $e^-e^+ \rightarrow Z \gamma h$ in an effective Lagrangian framework. For this study the beam polarization facility at the ILC along with the typical center-of-mass energy of 500 GeV is considered. The reach of the ILC with an integrated luminosity of 300 fb$^{-1}$ in the determination of $CP$-conserving parameters is obtained. Sensitivity of the probe of each of these couplings in the presence of other couplings is investigated. The most influential coupling parameters are $\bar c_W=-\bar c_B$. Other parameters of significant effect are $\bar c_{HW}$ and $\bar c_{HB}$. A detailed study of the various kinematic distributions represents possibilities to disentangle the effect of some of these couplings.
\end{abstract} 
\maketitle
\newpage
\section{Introduction}
With the discovery of the new resonance of mass around 125 GeV by the ATLAS and CMS collaborations at the LHC in 2012, a change of paradigm has taken \cite{cms, atlas, Moriond1, Moriond2, Moriond3, Moriond4, Moriond5,Aad:2013wqa,Chatrchyan:2012ufa,Chatrchyan:2013lba,Aad:2012tfa,Khachatryan:2014ira,Aad:2014eha}. The discovery has unambiguously established the role of the Higgs mechanism in electroweak symmetry breaking (EWSB). All the properties of the new particle measured so far are consistent with that of the standard Higgs boson. Thus, one may be tempted to conclude that for all practical purposes, the newly found particle is like that of the Standard Model (SM) Higgs boson, and new physics effects are decoupled as far as the Higgs sector is concerned. At the same time, it is well known that there are difficulties associated with the Higgs sector of the SM that need to be addressed. The main difficulty is the hierarchy problem associated with the quadratically diverging quantum corrections to the mass of the Higgs boson when computed in the SM. There is no remedy to this difficulty within the SM, and for a Higgs boson of mass 125 GeV, the new physics effects should show up within the TeV range to cure this malady. Assuming that the new physics effects are expected to appear only indirectly in the Higgs sector, it is natural to consider these effects through effective couplings of the Higgs bosons, with itself as well as with the gauge bosons and heavy fermions. Precise measurement of these couplings is very essential to establish the true nature of the EWSB mechanism. While the LHC is capable of probing some of these couplings \cite{Gabrielli:2013era}, especially the Higgs couplings with the gauge bosons and top quark, one may need to rely on a cleaner machine like the International Linear Collider (ILC) \cite{ILC1,ILC2,polarizationreview,Boos:2014xza} for the required precision. Within an effective Lagrangian, the effect of new physics could be studied in the various couplings through the quantum corrections they acquire. Such an effective Lagrangian basically encodes the new physics effects in higher-dimensional operators with anomalous couplings. \\

The study of the Higgs sector through an effective Lagrangian goes back to Refs.\cite{Weinberg:1978kz,Weinberg:1980wa,Georgi:1994qn,Buchmuller:1985jz,Hagiwara:1993ck,Hagiwara:1993qt,Alam:1997nk,
genuined6,Giudice:2007fh,Contino2010a,Contino2010b,Grober2011,Grzadkowski:2010es}.
More recently, the Lagrangian including a complete set of dimension-six operators was studied by 
Refs. \cite{Baak:2012kk, Einhorn:2013kja,Contino:2013kra,Willenbrock:2014bja}. For some of the recent references discussing the constraints on the anomalous couplings within different approaches, please see Refs.
\cite{Bonnet:2011yx,Corbett:2012dm,Chang:2013cia,Elias-Miro:2013mua,Banerjee:2013apa,Boos:2013mqa,
Masso:2012eq,Han:2004az, Corbett:2012ja,Dumont:2013wma, Pomarol:2013zra,Ellis:2014dva, Belusca-Maito:2014dpa, Gupta:2014rxa}.
Reference \cite{Ellis:2014dva} studied the $h+V$, where $V= Z, W$ associated production at the LHC and Tevatron to discuss the bounds obtainable from the global fit to the presently available data, whereas Ref.~\cite{Belusca-Maito:2014dpa} has discussed the constraint on the parameters coming from the LHC results as well as other precision data from LEP, SLC, and Tevatron. 
Experimental studies on the Higgs couplings at the LHC are presented in, for example, Refs. \cite{ 1307.1427,Teyssier:2014hta}.\\

The measurement of trilinear Higgs couplings is best done through the process $e^+ e^-\rightarrow Zhh$  \cite{De Rujula:1991se,1,2,3,4,5,6,7,8,9,10}.  At the same time, this process also depends on the Higgs-gauge boson couplings, $ZZh$ and $ZZhh$, which will affect the determination of the $hhh$ coupling. Another process that could probe the $hhh$ couplings is $e^+e^- \rightarrow \nu_e \bar\nu_e hh$ following the $WW$ fusion \cite{4,5,6,7}, which is also affected by the $WWh$ and $WWhh$ couplings. In a recent study \cite{Kumar:2014zra}, we investigated the effect of the $VVh$ coupling, where $V=Z,~W$, in the extraction of the $hhh$ coupling, and found that a precise knowledge of the $WWh$ and $ZZh$ couplings is necessary to derive information regarding the trilinear couplings. Apart from this, an investigation of Higgs-gauge boson couplings has been performed in the recent study of WWh production at ILC \cite{Kumar:2015eea}. The anomalous couplings like $hZ\gamma$ and $h\gamma \gamma$ has been studied \cite{Cao:2015fra, Ren:2015uka} in the process of $e^- e^+ \rightarrow h \gamma$. The Table \ref{table:constraints} contains obtained limits on coupling parameters in previous studies.\\

The process $e^+ e^-\rightarrow Z \gamma h$ is well suited to study the Higgs to neutral gauge boson couplings \cite{De Rujula:1991se,1,2,3,4,5,6,7,8,9,10}. This process is influenced by the trilinear couplings like $Z\gamma h,~ h\gamma\gamma$ and $ZZh\gamma$ which can contaminate the effects of $ZZh$ couplings. In this paper we will focus our attention on this process in some detail within the framework of the effective Lagrangian. One goal of this study is to investigate CP-conserving Higgs sector through Higgs to gauge boson couplings and to understand the significance of other couplings in their measurement.

\begin{table}[H]
\begin{center}
\begin{tabular}{cccc}
\hline
~~~~Coefficients ~~~~~& ~~~~~Current Bound ~~~~~~&~~~~~~~~~~~$\sigma(\gamma h)$ ~~~~~~~~~~~~& ~~~~~~~~~~~~~$\sigma(Zh)$~~~~~~~~~ \\ \cline{1-4}
&&&\\[-3mm]
$\bar c_{HW}$               &[-0.042, 0.008]                    &[-0.0050, 0.0033]  & [-1.8, 1.8]$\times 10^{-4}$  \\
$\bar c_{HB}$                &[-0.053, 0.044]                    &[-0.0033, 0.0050]  & [-1.8, 1.8]$\times 10^{-4}$  \\
$\bar c_{\gamma}$        &[-4.0, 2.3]$\times 10^{-4}$  &[-0.0012, 0.0028]  & [-9.0,9.0]$\times 10^{-4}$  \\[-3mm]
&&&\\
\hline
\end{tabular}
\vspace{2mm}
\caption{Current 95\% CL bounds (2nd column) and future Higgs factory 2$\sigma$ exclusion sensitivities (3rd and 4th columns) on the coefficients of dim.-6 operators that contribute to $e^+ e^-\rightarrow \gamma h$.}
\label{table:constraints}
\end{center}
\end{table}

The paper is presented in the following way. In Sec.~\ref{sec:setup}, the effective Lagrangian will be presented, with the currently available constraint on the parameters. In Sec.~\ref{sec:discussions}, the process under consideration will be presented, with details. In Sec.~\ref{sec:summary}, the results will be summarized. 

\section{General Setup}\label{sec:setup}
 References \cite{Giudice:2007fh,Contino2010a,Contino2010b,Grober2011,Contino:2013kra,Ellis:2014dva, HEL} present the most general effective Lagrangian with dimension-six operators involving the Higgs bosons. Part of this Lagrangian relevant to the process $e^+ e^-\rightarrow Z \gamma h$ considered in this paper is given by

\begin{eqnarray}
{\cal L}_{\rm Higgs}^{\rm anom} &=& \frac{\bar c_T}{2 v^2}(\Phi^\dag {\overleftrightarrow D}^\mu \Phi) (\Phi^\dag {\overleftrightarrow D}_\mu \Phi)+\frac{\bar c_{\gamma}}{m_W^2} g'^2~\Phi^\dag \Phi B_{\mu\nu} B^{\mu\nu}  \nonumber \\
&& + \frac{\bar c_{HW}}{m_W^2} ig~\big(D^\mu \Phi^\dag \sigma_{k} D^\nu \Phi\big) W_{\mu \nu}^k + \frac{\bar c_{HB}}{m_W^2} ig'~ \big(D^\mu \Phi^\dag D^\nu \Phi\big) B_{\mu \nu} \nonumber\\
&&+ \frac{\bar c_{W}}{2m_W^2} ig~\big( \Phi^\dag \sigma_{k} \overleftrightarrow{D}^\mu \Phi \big)  D^\nu  W_{\mu \nu}^k +\frac{\bar c_{B}}{2 m_W^2} ig'~\big(\Phi^\dag \overleftrightarrow{D}^\mu \Phi \big) \partial^\nu  B_{\mu \nu},
\label{eq:Leff}
\end{eqnarray}
where
\(
  \Phi^\dag {\overleftrightarrow D}_\mu \Phi = 
    \Phi^\dag (D_\mu \Phi) - (D_\mu\Phi^\dag) \Phi \ ,
\) $D_\mu$ being the appropriate covariant derivative operator, and $\Phi$, the usual Higgs doublet in the SM. Also, $W_{\mu\nu}^k$ and $B_{\mu\nu}$ are the field tensors corresponding to the $SU(2)_L$ and $U(1)_Y$ of the SM gauge groups, respectively, with gauge couplings $g$ and $g'$, in that order. $\sigma_k$ are the Pauli matrices, and $\lambda$ is the usual (SM) quartic coupling constant of the Higgs field.
The above Lagrangian, leads to the following in the unitary gauge and mass basis
\cite{HEL}
\begin{eqnarray}
  {\cal L}_{h,Z, \gamma} ^{\rm anom}=&& -\frac{1}{4} g_{hZZ}^{(1)} Z_{\mu\nu} Z^{\mu\nu} h -
 g_{hZZ}^{(2)} Z_{\nu}\partial_{\mu} Z^{\mu\nu} h \nonumber\\
 &&+\frac{1}{2} g_{hZZ}^{(3)} Z_{\mu} Z^{\mu} h -\frac{1}{2} g_{h\gamma Z}^{(1)} Z_{\mu\nu} F^{\mu\nu} h - g_{h \gamma Z}^{(2)}Z_{\nu} \partial_{\mu} F^{\mu\nu} h \nonumber\\ 
 && - \frac{1}{4} g_{hh\gamma Z}^{(1)} Z_{\mu\nu} F^{\mu\nu} h^2 - \frac{1}{2} g_{hh\gamma Z}^{(2)}Z_{\nu} \partial_{\mu} F^{\mu\nu} h^2 
\label{eq:LagPhys}
\end{eqnarray}

\begin{table}[H]
\begin{tabular}{ll}
\hline 
\hline 
&\\
Trilinear and quartic CP-conserving couplings\\[2mm]
$g_{hzz}^{(1)}=\frac{2g}{\cos\theta_W^2 m_W} \left[ \bar{c}_{HB} \sin\theta_W^2 - 4\bar{c}_{\gamma} \sin\theta_W^4 + c_W^2 \bar{c}_{HW}  \right]$,&\\[3mm]
$g_{hzz}^{(2)} =\frac{g}{\cos\theta_W^2 m_W} \left[ (\bar{c}_{HW}+ \bar{c}_W) \cos\theta_W^2 + (\bar{c}_B + \bar{c}_{HB}) \sin\theta_W^2  \right]$, &\hspace{29mm} $g_{hzz}^{(3)}  = \frac{g m_Z}{\cos\theta_W} \left[ 1-2 \bar{c}_T  \right]$\\ [3mm]
$g_{h\gamma z}^{(1)}  = \frac{g \sin\theta_W}{\cos\theta_W m_W} \left[ \bar{c}_{HW} -\bar{c}_{HB} + 8\bar{c}_{\gamma} \sin\theta_W^2 \right]$,&~~
$g_{h\gamma z}^{(2)}  = \frac{g \sin\theta_W}{\cos\theta_W m_W} \left[ \bar{c}_{HW} -\bar{c}_{HB} -\bar{c}_B + \bar{c}_W \right]$\\ [3mm]
$g_{hh\gamma z}^{(1)}  = \frac{g^2 \sin\theta_W}{2 \cos\theta_W m_W^2} \left[ \bar{c}_{HW} -\bar{c}_{HB} + 8\bar{c}_{\gamma} \sin\theta_W^2 \right]$,&
$g_{hh\gamma z}^{(2)}  = \frac{g^2 \sin\theta_W}{2 \cos\theta_W m_W^2} \left[ \bar{c}_{HW} -\bar{c}_{HB} -\bar{c}_B + \bar{c}_W \right]$\\
&\\
\hline
\hline
\end{tabular}
\caption{Physical couplings in Eqs. (\ref{eq:LagPhys}) are given in terms of the effective couplings in Eq. (\ref{eq:Leff}), where $\theta_W$ being the weak mixing angle.}
\label{table:couplings}
\end{table}

The physical couplings relevant to the process $e^+e^-\rightarrow Z\gamma h$, and associated with the Lagrangian in Eqs.~(\ref{eq:LagPhys}) expressed in terms of the effective couplings presented in Eq.~(\ref{eq:Leff}) are listed in Table~\ref{table:couplings}.
In total, there are five parameters which are relevant to the process considered, viz,  $\bar{c}_T, \bar{c}_{\gamma}, \bar{c}_B, \bar{c}_W, \bar{c}_{HB}, \bar{c}_{HW}$.  
These anomalous coefficients $\bar{c}_T, \bar{c}_{HW}, \bar{c}_{HB}, \bar{c}_{\gamma}$ are expected to be of the order
\begin{equation}
 \bar{c}_T \sim \mathcal{O}\left( \frac{g_{\sc NP}^2 v^2}{M^2}\right) ~~{\rm and}~~ 
 \bar{c}_{HW}, \bar{c}_{HB}, \bar{c}_{\gamma} \sim \mathcal{O}\left( \frac{{g_{\sc NP}^2} M_{W}^2}{16 \pi^2 M^2}\right),
\end{equation}
where $g_{\sc NP}$ denotes the generic coupling of the new physics, and $M$ is the new physics scale. This indicates that
these couplings can be significantly large for strongly coupled physics. In contrast the coefficients of the operators
such as $\bar{c}_W$ and $\bar{c}_B$ are given by
\begin{equation}
 \bar{c}_B, \bar{c}_W \sim \mathcal{O}\left( \frac{m_W^2}{M^2}\right)
\end{equation}
and therefore, expected to be relatively suppressed or enhanced according to the ratio ${g/g_{\sc NP}}$. \\

Coming to the experimental bounds, electroweak precision data put the following constraints~\cite{Baak:2012kk},
\begin{equation}
 \bar{c}_T(m_Z) \in [-1.5, 2.2] \times 10^{-3} ~{\rm and}~ (\bar{c}_W(m_Z) + \bar{c}_B(m_Z))\in [-1.4, 1.9]\times 10^{-3}
 \label{eq:cTcWcBconst}
\end{equation}
This means we can safely ignore the effect of $\bar c_T$ in our analysis. On the other hand, $\bar{c}_W$ and $\bar{c}_B$ are not independently constrained, leaving the possibility of having  large values with a cancellation between them as per the above constraint. $\bar c_W$ itself along with $\bar c_{HW}$ and $\bar c_{HB}$ are constrained from LHC observations on the associated production of the Higgs along with $W$ in Ref. ~\cite{Ellis:2014dva}.
Considering the Higgs-associated production along with $W$, ATLAS and CMS along with D0 put a limit of  \( \bar{c}_W \in \big[-0.03, 0.01\big] \), when all other parameters were set to zero. A global fit using various information from ATLAS and CMS including signal-strength information constrains the region in the $\bar c_W-\bar c_{HW}$ plane, leading to a slightly more relaxed limit on $\bar c_W$ and a limit of about \( \bar{c}_{HW} \in \big[-0.04, 0.01\big] \). The limit on $\bar{c}_{HB}$ estimated using a global fit in Ref.~\cite{Ellis:2014dva} is about \(\bar{c}_{HB} \in [-0.05, 0.05]\), while a limit of about \(\bar{c}_{\gamma}\in [-0.04, 0.03]\) with a one-parameter fit.  

The purpose of this study is to understand how to exploit a precision machine like the ILC to investigate a suitable process so as to derive information regarding these couplings. In the next section, we shall explain the process of interest in the present case and discuss the details to understand the influence of one or more of the couplings mentioned above.

\section{Analyses of the process considered} \label{sec:discussions}
\begin{figure}[H] \centering
\begin{tabular}{c c}
\hspace{20mm}
\includegraphics[angle=0, width=80mm]{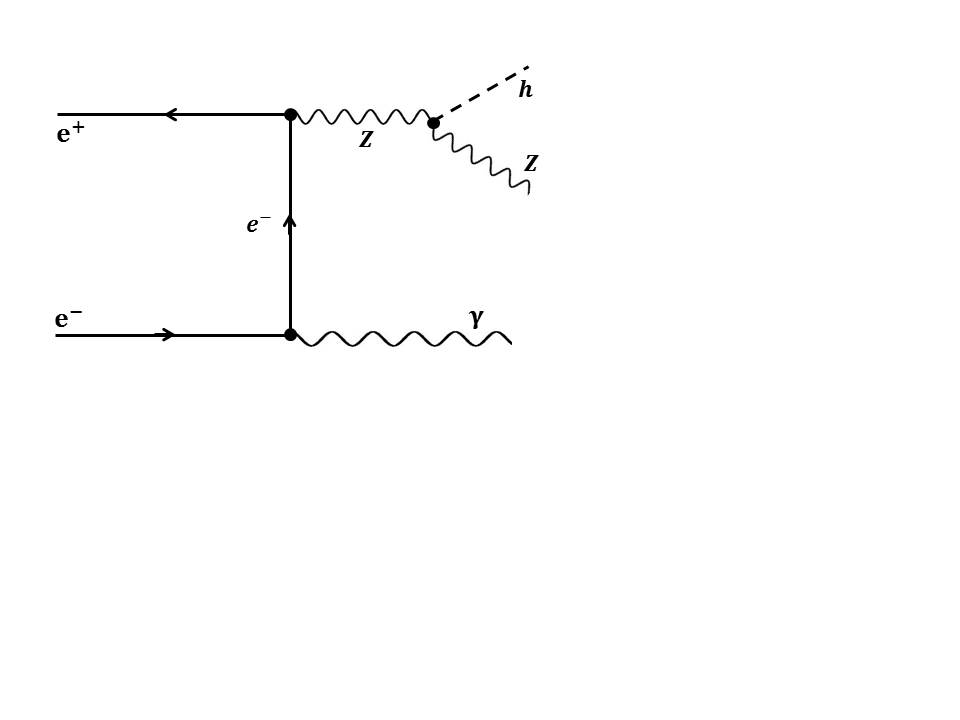} &
\vspace{-28mm}\hspace{-25mm}
\includegraphics[angle=0, width=80mm]{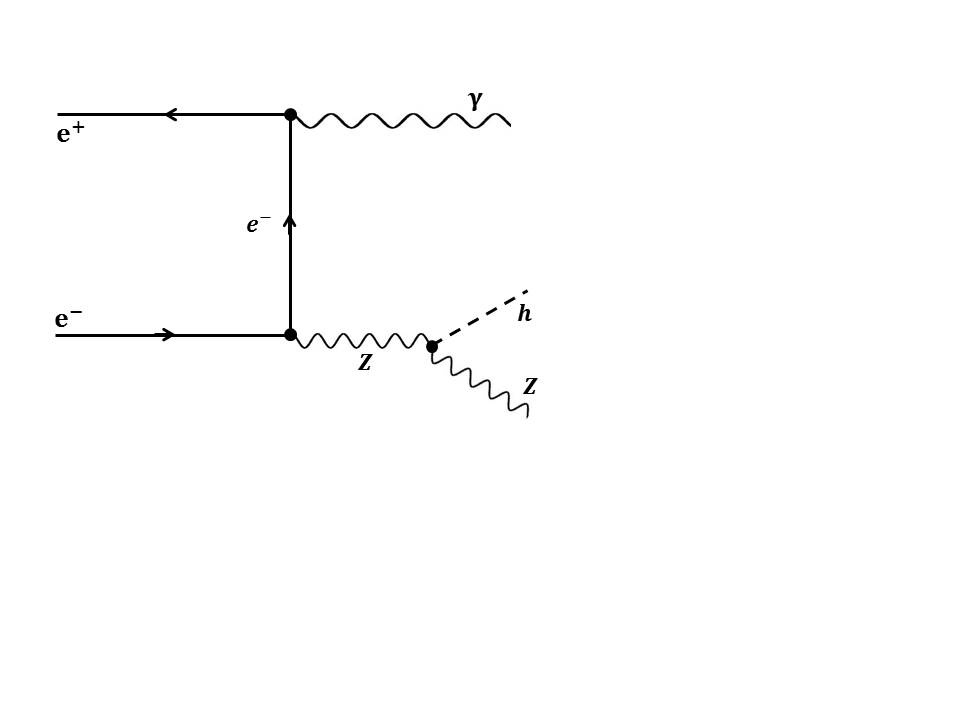} \\
\end{tabular}
\caption{Feynman diagrams contributing to the process $e^- e^+ \rightarrow Z \gamma h$ in the SM.}
\label{fig:feynmandiag1}
\end{figure}

The Feynman diagrams corresponding to the process $e^- e^+ \rightarrow Z \gamma h$  in the SM are given in Fig.\ref{fig:feynmandiag1}. This process is basically influenced by Higgs to neutral gauge boson couplings like $ZZh, ~ Z\gamma h$, $h \gamma \gamma$, and $ZZ \gamma h$ apart from the fermionic couplings, which are taken to be the standard couplings in our study. 

The effective Lagrangian, Eq. (\ref{eq:Leff}), apart from allowing the existing Higgs and gauge boson couplings to be nonstandard, introduces new couplings which are absent in the SM. In a specific model such effects appear at higher orders with a new particle present in the loops. When the masses of such particles are taken to be large, the effect of such quantum correction can be considered in terms of changed couplings. Such effective couplings arising in the present analysis are presented in Table~\ref{table:couplings}. \\
Our numerical analyses are carried out using {\sc madgraph} \cite{madgraph, Alwall:2014hca}, with the effective Lagrangian implemented through {\sc feynrules} \cite{feynrules, HEL}.

\begin{figure}[H] \centering
\begin{tabular}{c c}
\hspace{3mm}
\includegraphics[angle=0,width=80mm]{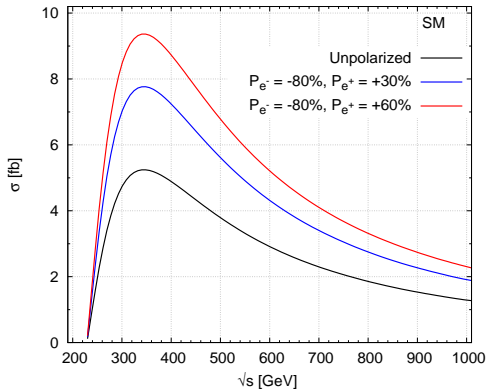} 
\end{tabular}
\vspace{-10mm}
\caption{ The total cross section against $\sqrt{s}$ in the SM with and without polarized beams.}
\label{fig:cs1}
\end{figure}
 
\begin{figure}[H] \centering
\begin{tabular}{c c}
\hspace{6mm}
\includegraphics[angle=0,width=80mm]{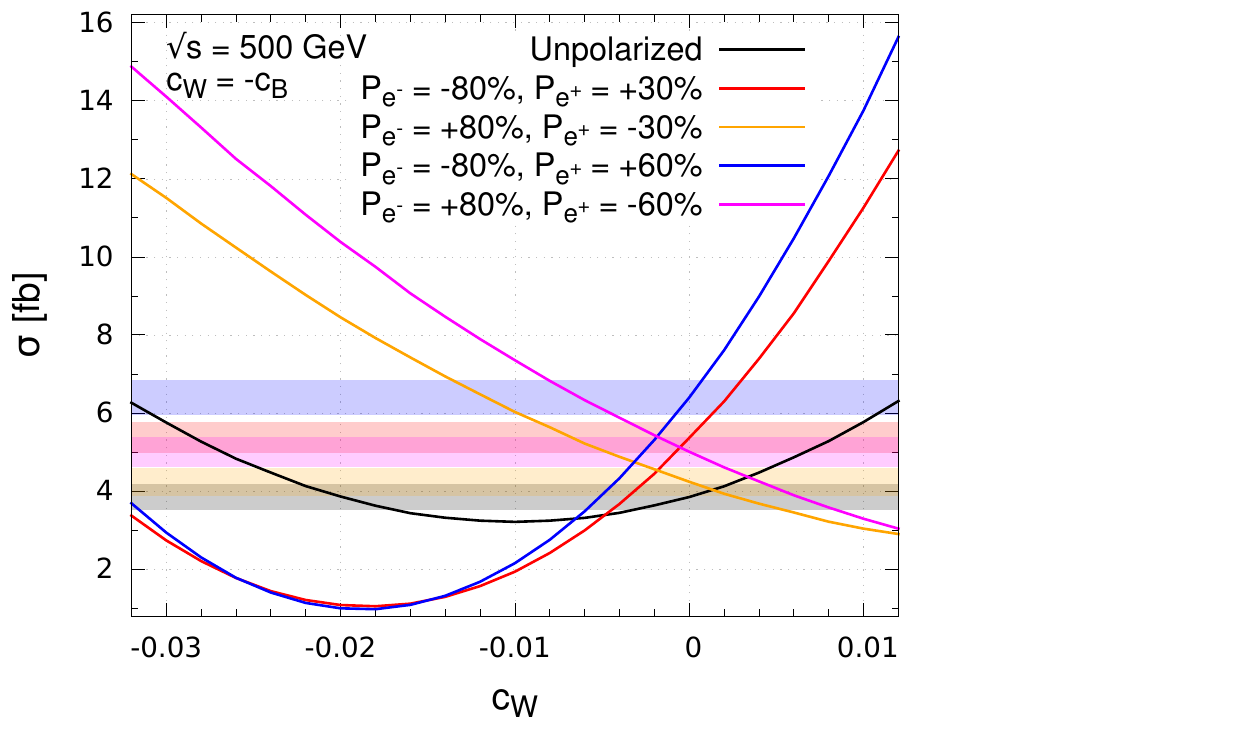} &
\hspace{-14mm}
\includegraphics[angle=0,width=80mm]{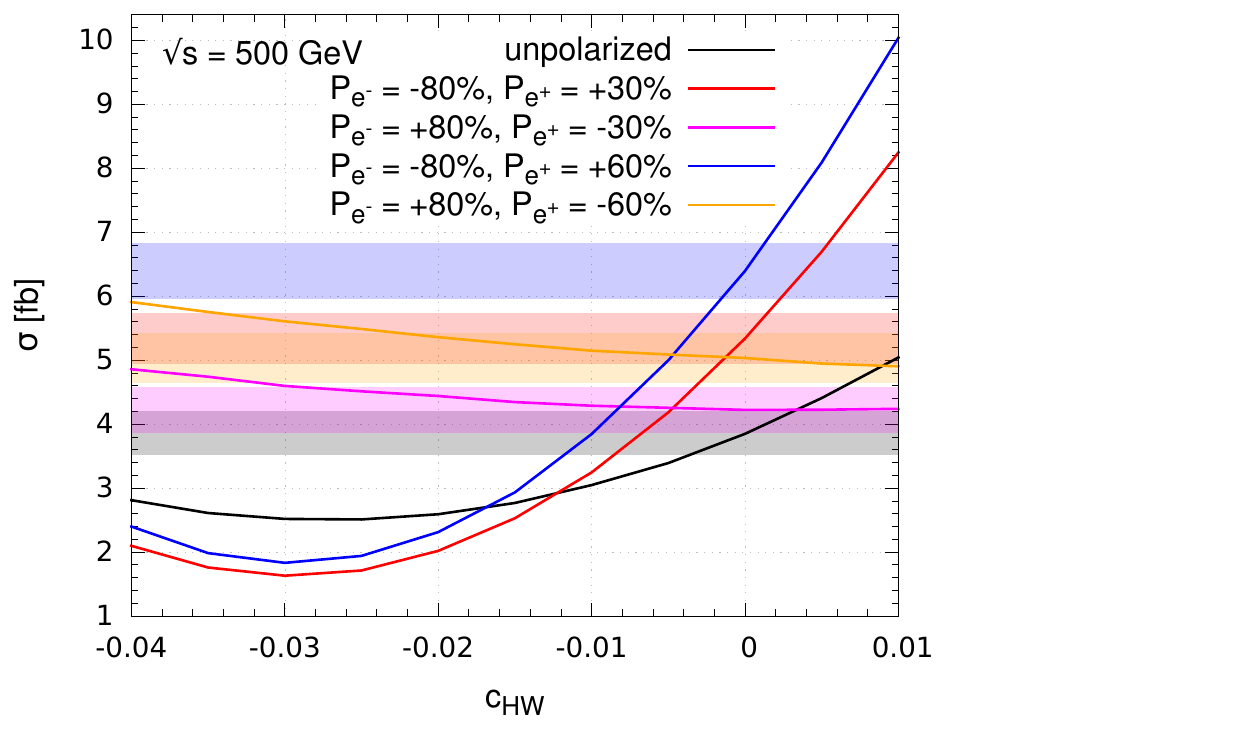} \\
\hspace{6mm}
\includegraphics[angle=0,width=80mm]{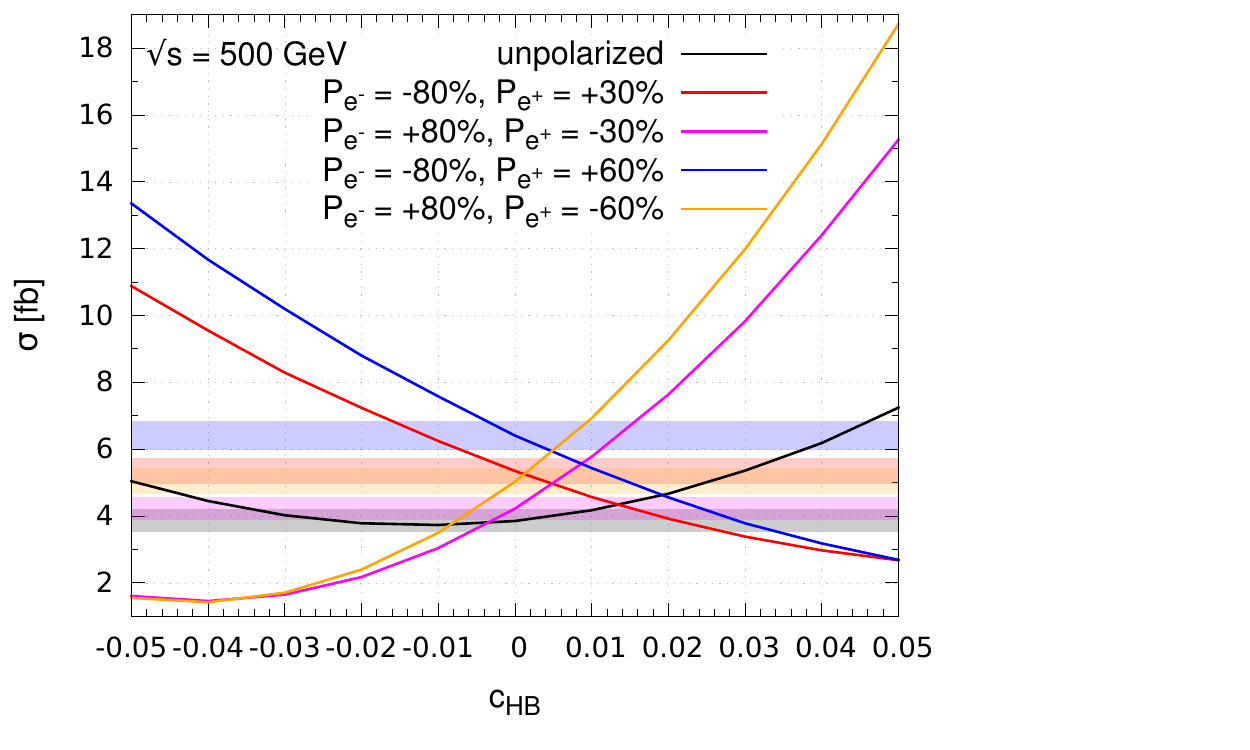} &
\hspace{-14mm}
\includegraphics[angle=0,width=80mm]{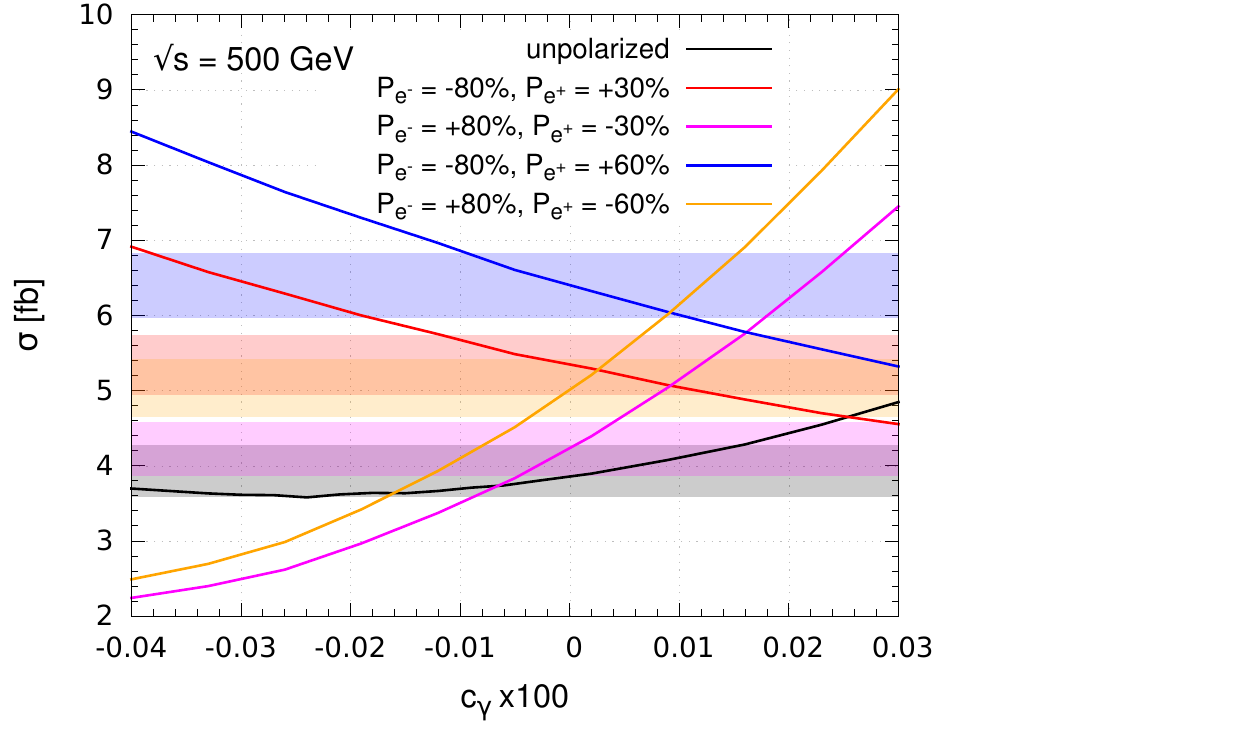} \\
\end{tabular}
\vspace{-4mm}
\caption{The total cross section against anomalous coupling parameters at $\sqrt{s}=500$ GeV, where the gray, red and blue bands correspond to $3\sigma$ deviations from the SM with unpolarized and polarized beams with an integrated luminosity of 300 fb$^{-1}$, respectively.Note that here, as well as in all figures henceforth, we have removed the ``bar" from the symbols denoting the CP-conserving parameters for convenience.}
\label{fig:cs2}
\end{figure}

As the first observable, we consider the cross section$^{1}$\footnotetext[1]{The most general formula is available in the Appendix-A.}. Figure \ref{fig:cs1} presents the total cross section against the center-of-mass energy for the $Z\gamma h$ production. The cross section peaks around the center-of-mass energy of $350$ GeV with a value of about 5.2 fb, which slides down to about 3.8 fb at 500 GeV. In order to avoid any complications arising from the threshold effects, we perform our analysis for an ILC running at a center-of-mass energy of 500 GeV, sufficiently away from the threshold value. This is one of the planned center-of-mass energy of the proposed ILC. As expected, the polarization hugely improves the situation. The case of a typical polarization combination expected at the ILC, 80\% left-polarized electron beam and 30\% right-polarized positron beam, is considered \cite{polarizationreview}, along with the case of an 80\% left-polarized electron beam and a 60\% right-polarized positron beams, which are expected in the upgraded version of the ILC. In Fig.\ref{fig:cs2} the cross section against anomalous couplings parameters, $\bar c_W$ or $\bar c_{HW}$ or $\bar c_{HB}$ or $\bar c_{\gamma}$ at fixed center-of-mass energy of 500 GeV is considered along with the role of the polarized beams. In order to be consistent with the experimental constraint [Eq.~(\ref{eq:cTcWcBconst}], we choose $\bar c_B=-\bar c_W$ throughout our analysis, showing the high sensitivity of the cross section on this parameter. Assuming that no other couplings affect the process, the single parameter reach corresponding to the $3\sigma$ limit with 300~fb$^{-1}$ integrated luminosity is given in Table~\ref{table:constraints_cW}, while two parameter reach can be seen in Table~\ref{table:constraints_cHW}. The obtained limit in the case of unpolarized beam, which is improved with an 80\% left-polarized electron beam and a 30\% right-polarized positron beam. While the case with an 80\% left-polarized electron beam and a 60\% right-polarized positron beam does not change this limit significantly, the cross section is increased from about 5.4 fb to about 6.4 fb, enhancing the statistics. In our further analysis, we consider the baseline expectation of an 80\% left-polarized electron beam and a 30\% right-polarized positron beam.

\begin{table}[H]
\begin{center}
\begin{tabular}{|c|c|c|c|}
\hline
Couplings &Unpolarized &$e^- =-80\%, e^+ =+30\%$ &$e^- =-80\%, e^+ =+60\%$ \\ \cline{1-4}
&&&\\[-3mm]
                          &[-0.022, -0.017]  &  ------------        &------------    \\
$\bar c_{W}=-\bar c_{B}$  &[-0.003, +0.002]  &[-0.001, +0.001]  &[-0.001, +0.0007]  \\[2mm]
\hline
&&&\\[-3mm]
$\bar c_{HW}$             &[-0.004, +0.003]  &[-0.0019, +0.0017]  &[-0.0018, +0.0016]  \\
$\bar c_{HB}$             &[-0.035, +0.010]  &[-0.005, +0.005]  &[-0.004, +0.004]  \\
$\bar c_{\gamma}$         &[-0.026, +0.014]  &[-0.011, +0.014]  &[-0.009, +0.012]  \\[-3mm]
&&&\\
\hline
\end{tabular}
\vspace{2mm}
\caption{Showing the single parameter reach corresponding to the $3\sigma$ limit with an integrated luminosity of 300~fb$^{-1}$ at center-of-mass energy of 500 GeV, assuming that no other couplings affect the process.}
\label{table:constraints_cW}
\end{center}
\end{table}

\begin{figure}[H] \centering
\begin{tabular}{c c}
\hspace{-1mm}
\includegraphics[angle=0,width=80mm]{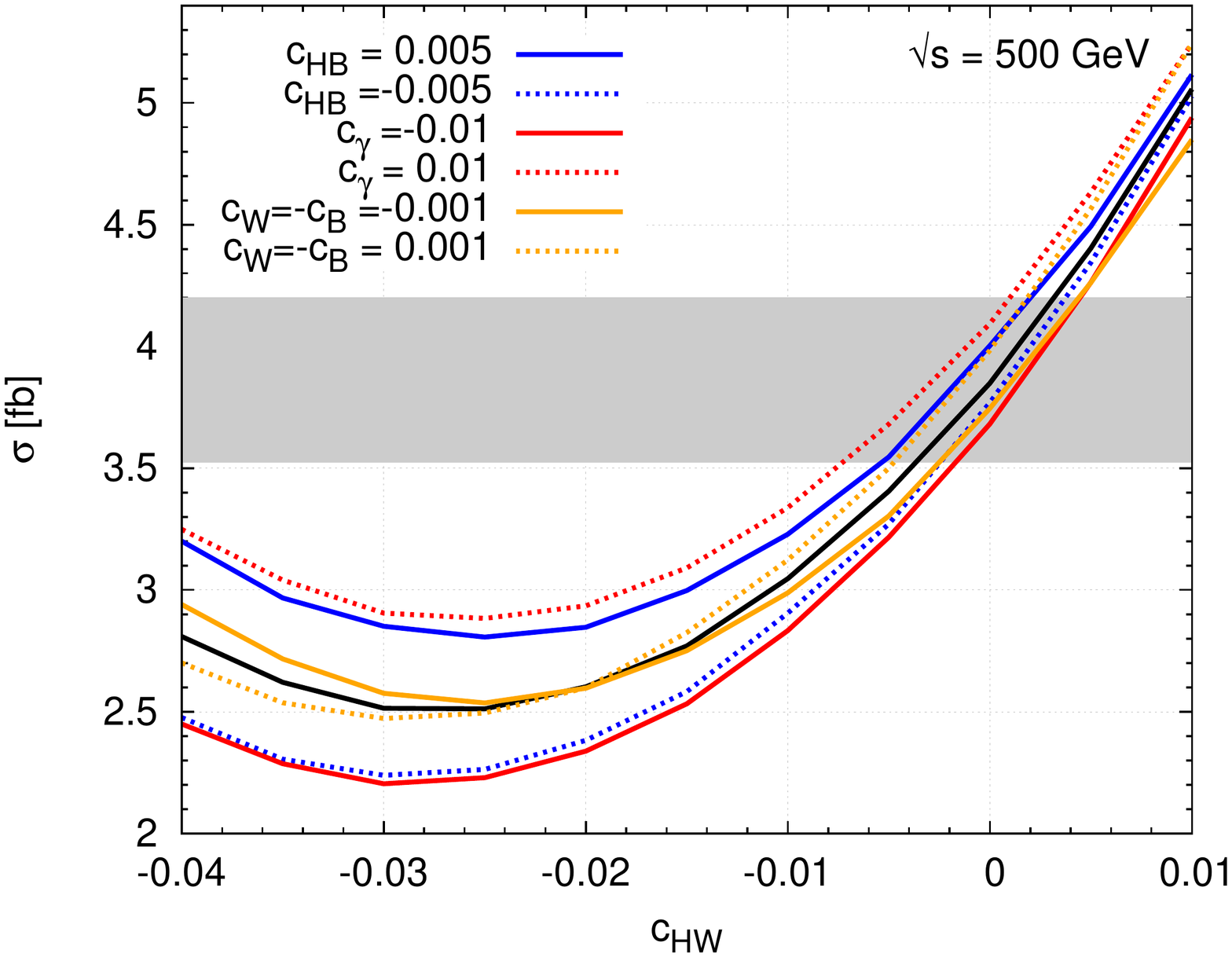} &
\vspace{-8mm}\hspace{-10mm}
\includegraphics[angle=0,width=80mm]{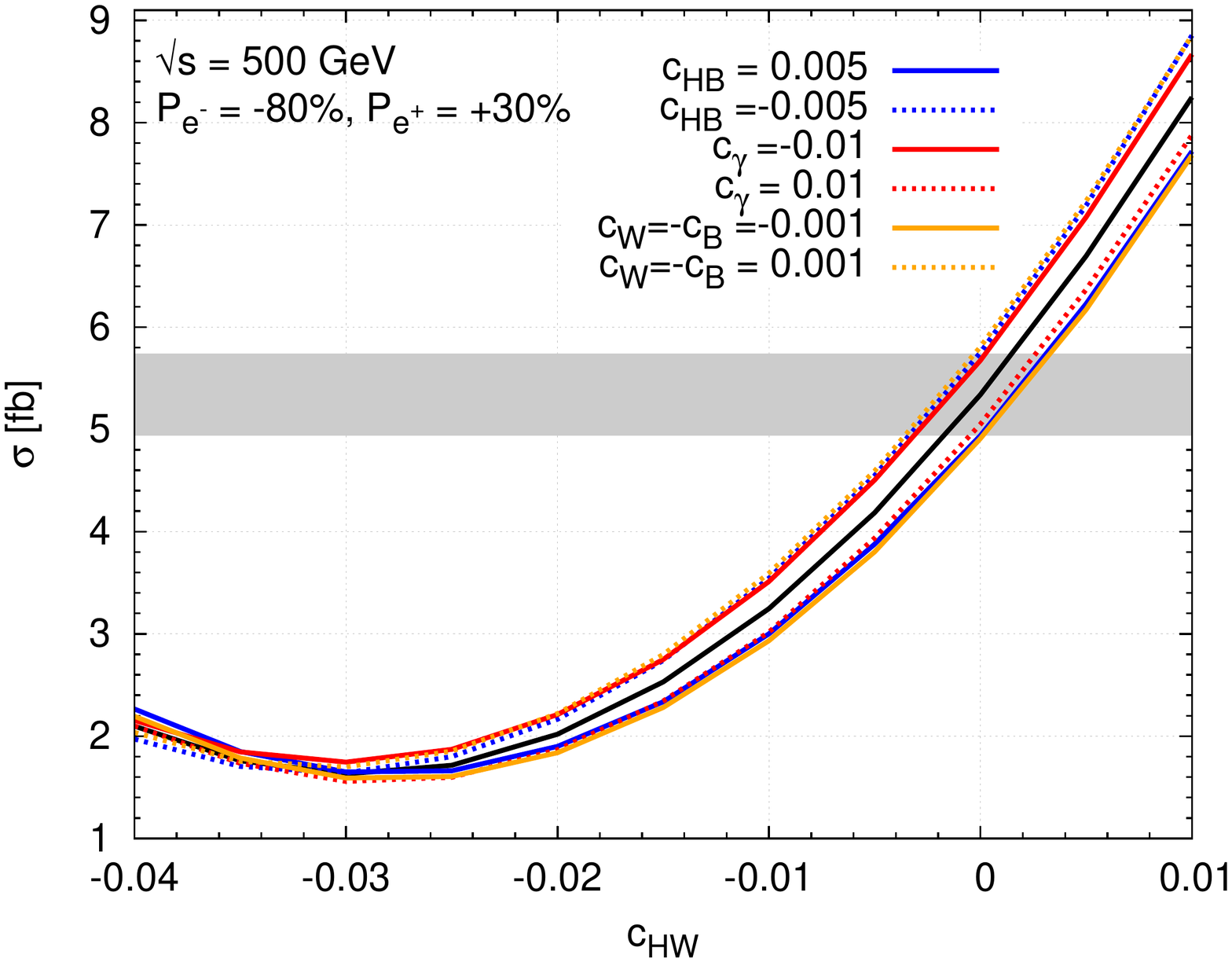} \\
\hspace{-1mm}
\includegraphics[angle=0,width=80mm]{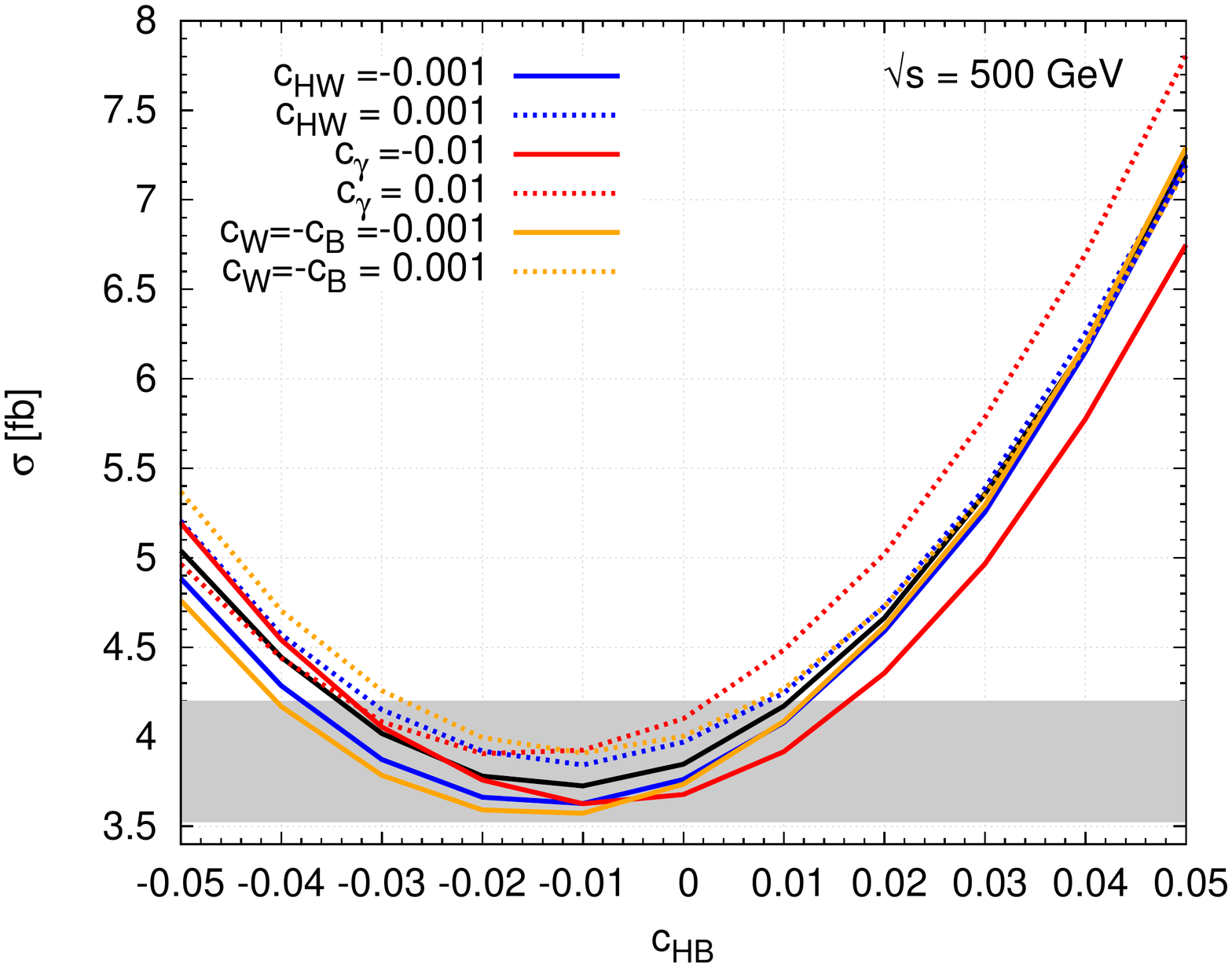} &
\vspace{-8mm}\hspace{-10mm}
\includegraphics[angle=0,width=80mm]{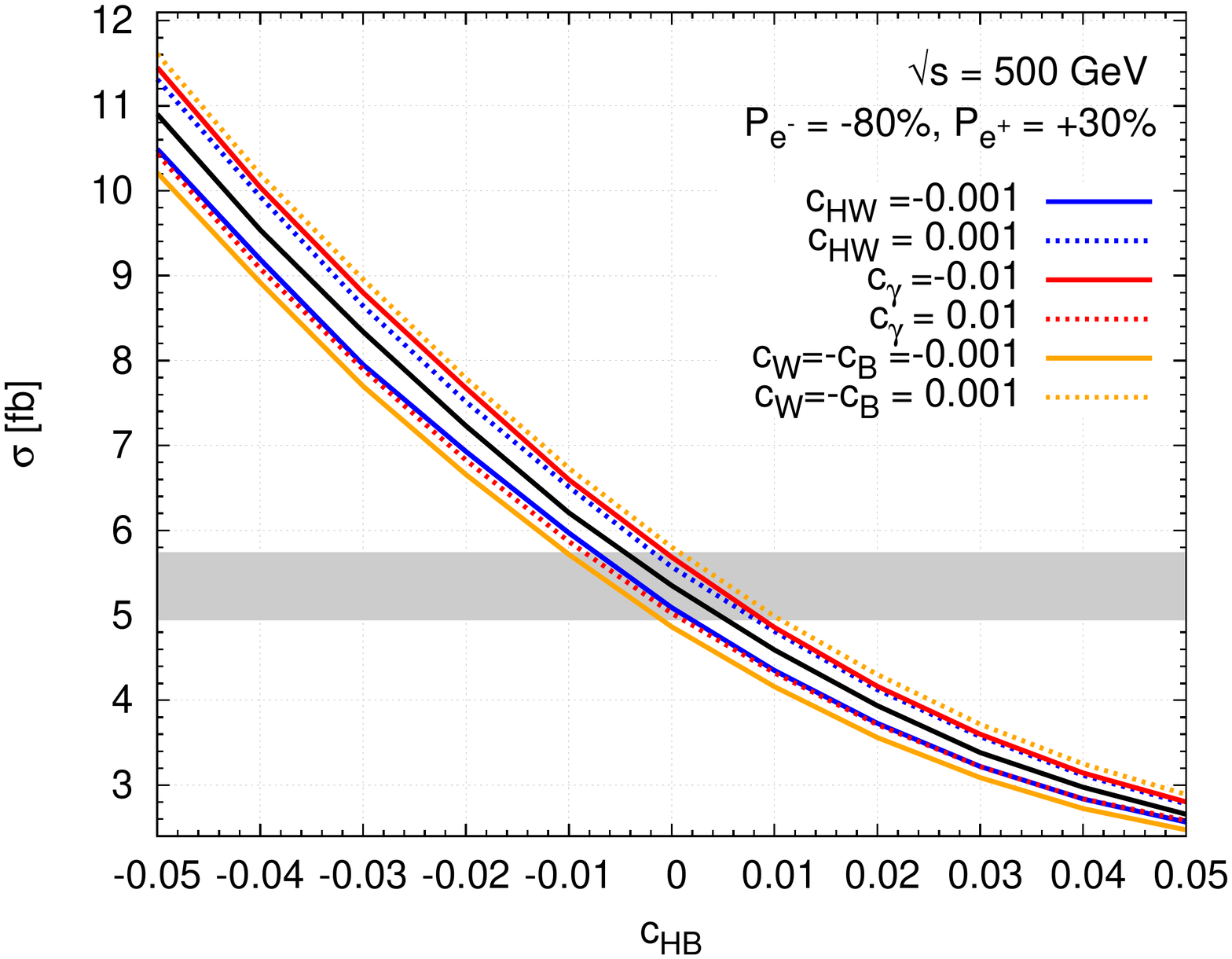} \\
\hspace{-1mm}
\includegraphics[angle=0,width=80mm]{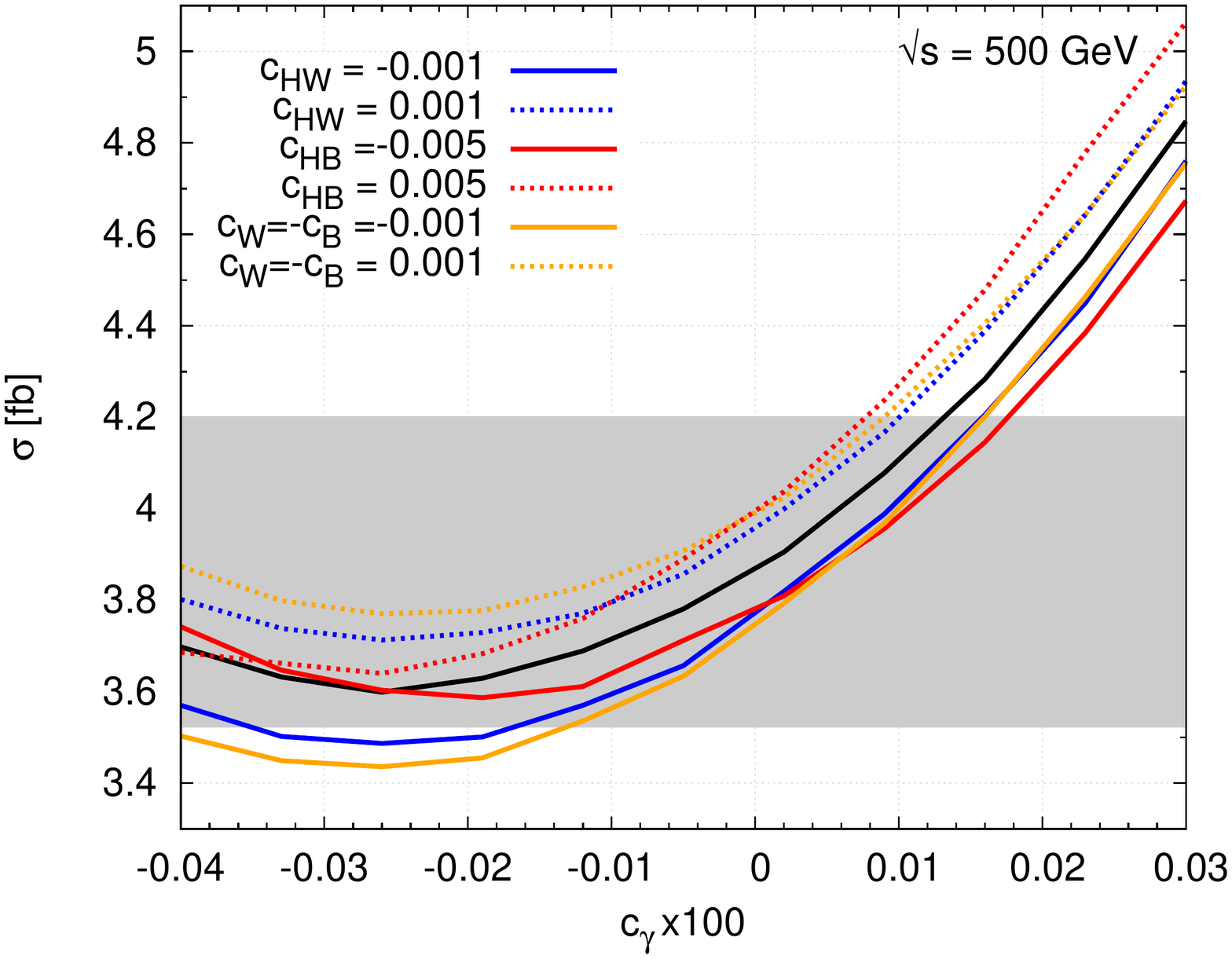} &
\hspace{-10mm}
\includegraphics[angle=0,width=80mm]{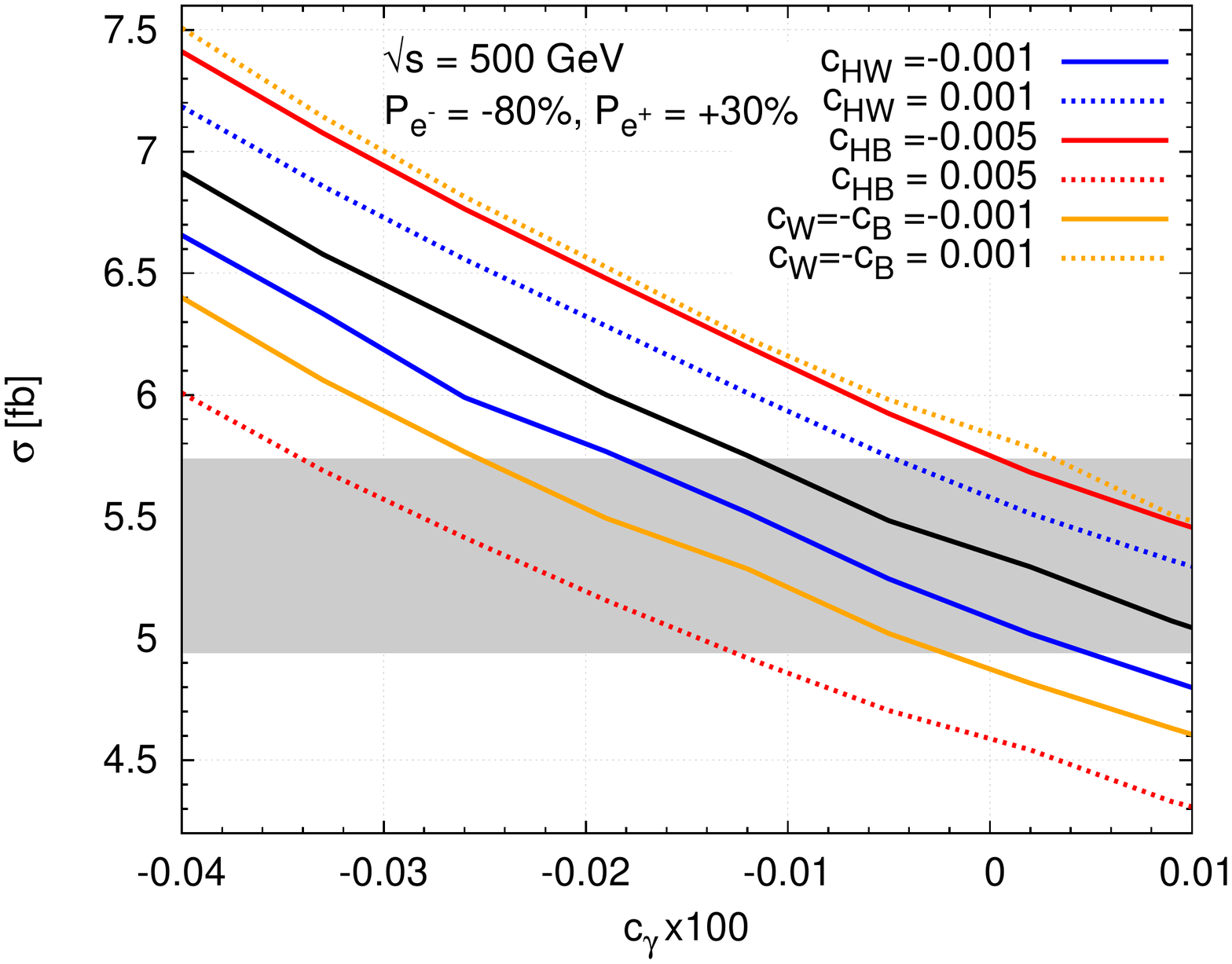}\\
\end{tabular}
\vspace{-10mm}
\caption{Cross section against, $\bar c_{HW}$ (top), $\bar c_{HB}$ (middle) and $\bar c_{\gamma}$ (bottom). The black solid line corresponds to the case when only $\bar c_{HW}$ or $\bar c_{HB}$ or $\bar c_{\gamma}$ is present. The center-of-mass energy is assumed to be $\sqrt{s}=500$ GeV. In each case, all other parameters are set to zero. The gray band indicates the $3\sigma$ limit of the SM cross section with unpolarized (left) and polarized beams (rigth) with an integrated luminosity of 300 fb$^{-1}$.}
\label{fig:cstcHW}
\end{figure}
 
Coming to the Fig.\ref{fig:cs2} \& \ref{fig:cstcHW}, here we present the single parameters analysis in the absence of other anomalous coupling parameters. Clearly, it can be seen that how the obtained limits are affected by the presence of other parameters. The gray band indicates the $3\sigma$ limit$^2$\footnotetext[2]{The formula can be seen in the Appendix-B.} of the SM cross section with unpolarized (left) and polarized beams (right) with an integrated luminosity of 300 fb$^{-1}$.   

\begin{table}[H]
\begin{center}
\begin{tabular}{|c|c|c|c|}
\hline
Couplings            &No other couplings  &$\bar c_{W}=-\bar c_{B}=-0.03$  &$\bar c_{W}=-\bar c_{B}=+0.01$ \\ \cline{1-4}
&&&\\[-3mm]
$\bar c_{HW}$        &[-0.080, +0.000]  &[-0.065, +0.055]  &[-0.105, +0.000]  \\
$\bar c_{HB}$        &[-0.100, +0.020]  &[-0.016, +0.010]  &[-0.011, +0.050]  \\[2mm]
\hline
&&&\\[-3mm]
$\bar c_{\gamma}$    &[-0.065, +0.020]  &[-0.175, +0.095]  &[-0.060, +0.035]  \\
$\bar c_{HB}$        &[-0.035, +0.025]  &[-0.150, +0.032]  &[-0.020, +0.045]  \\[2mm]
&&&\\[-3mm]\hline
Couplings            &No other couplings  &$\bar c_{HW}= -0.04$  &$\bar c_{HW}= +0.01$ \\ \cline{1-4}
&&&\\[-3mm]
$\bar c_{W}$         &[-0.070, +0.003]  &[-0.035, +0.025]  &[-0.082, +0.002]  \\
$\bar c_{HB}$        &[-0.170, +0.015]  &[-0.135, +0.030]  &[-0.200, +0.025]  \\[-3mm]
&&&\\
\hline
\end{tabular}
\caption{Showing the two parameters reach correspond to the $3\sigma$ limit with an integrated luminosity of 300~fb$^{-1}$ at center-of-mass energy of 500 GeV.}
\label{table:constraints_cHW}
\end{center}
\end{table}

\begin{figure}[H] \centering
\begin{tabular}{ccc}
\hspace{-3.5mm}
\includegraphics[angle=0,width=74mm]{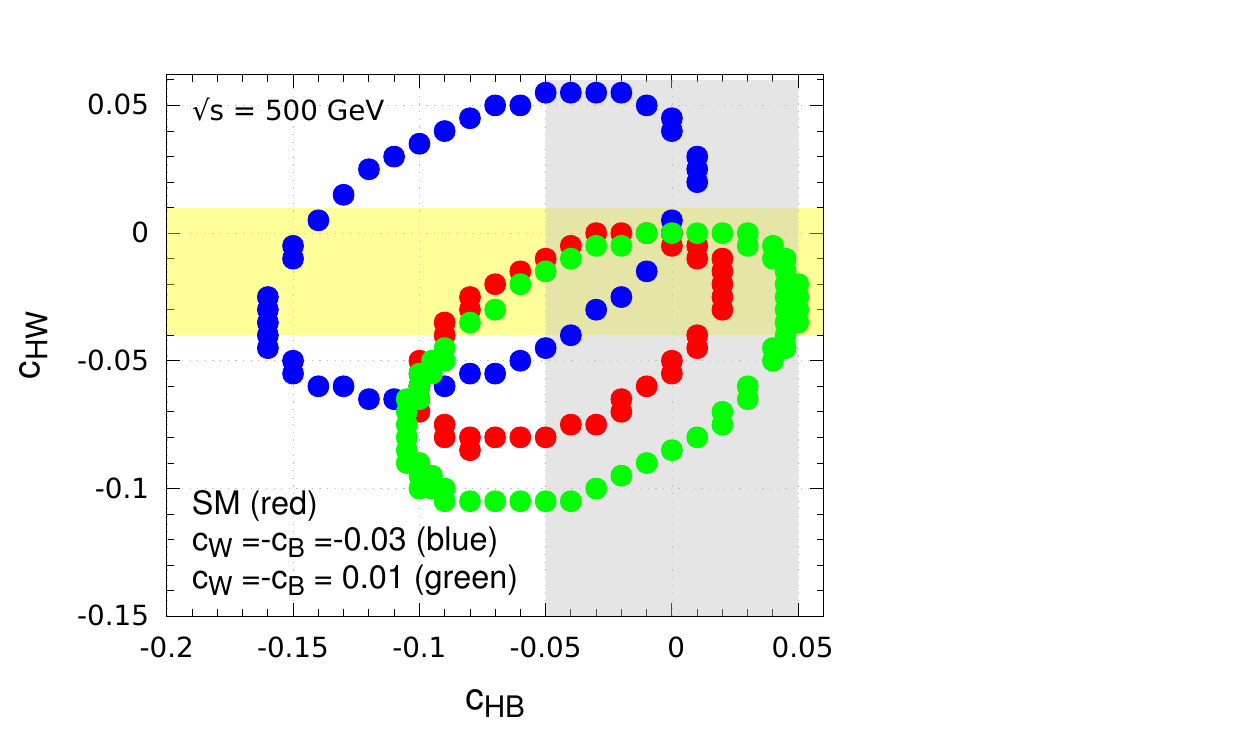} & 
\hspace{-25mm}
\includegraphics[angle=0,width=68mm]{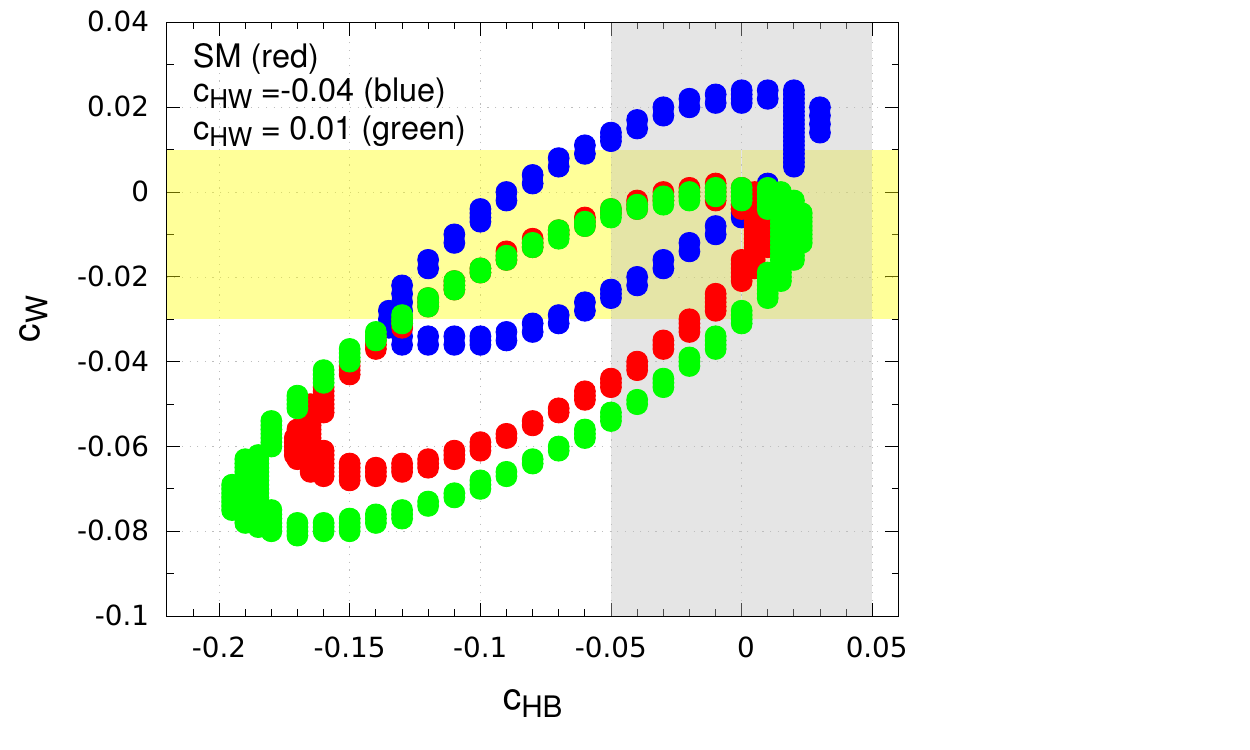}  &
\hspace{-20mm}
\includegraphics[angle=0,width=68mm]{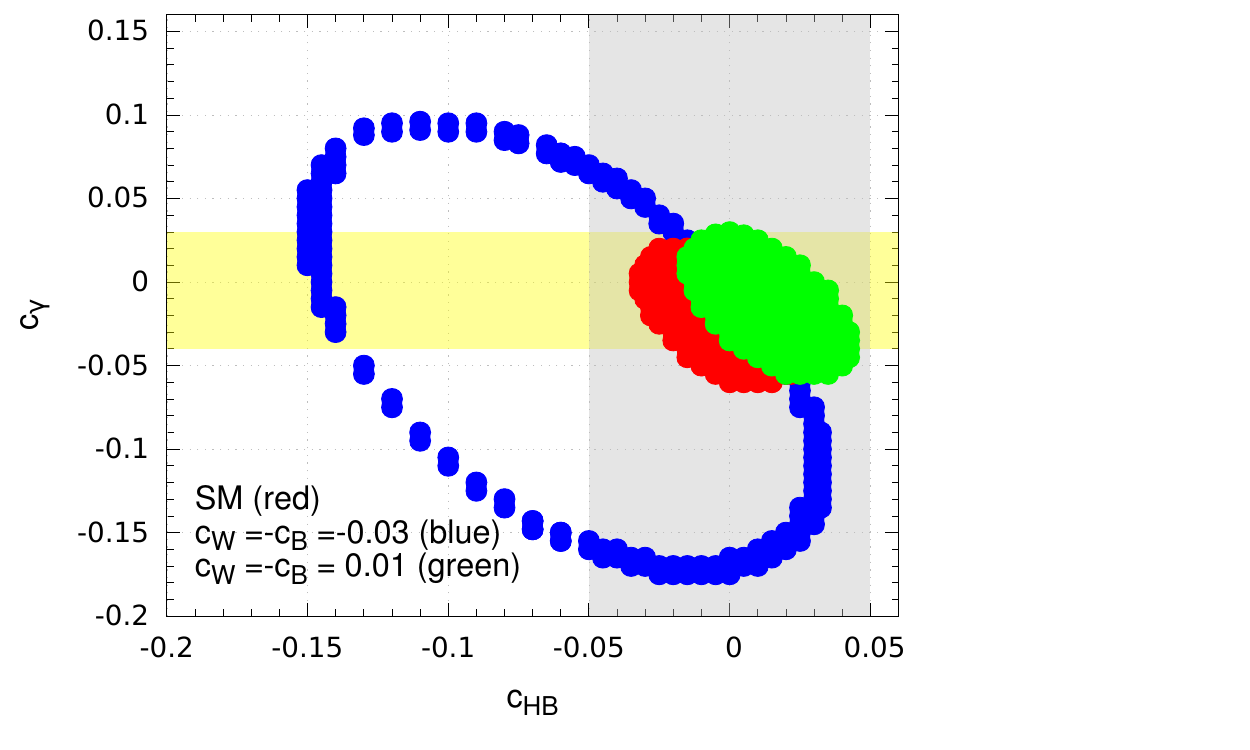} \\
\end{tabular}
\vspace{-4mm}
\caption{The ellipses correspond to regions in the $\bar c_{HB} -~\bar c_{HW}$ (top), $\bar c_{HB} -~\bar c_{W}$ (left, bottom) and $\bar c_{HB} -~\bar c_{\gamma}$ (right, bottom) planes with the total cross section within the $3\sigma$ limit of the SM cross section (red), and cross sections with $\bar c_W=-0.03$ (blue) and $\bar c_W=+0.01$ (green). An integrated luminosity of 300~fb$^{-1}$ is considered, and the center-of-mass energy is taken as 500 GeV. The yellow and gray bands correspond to the present limits of $\bar c_{W}$, $\bar c_{HW}$ and $\bar c_{HB}$, respectively.}
\label{fig:correlation1}
\end{figure}

 Fig.~\ref{fig:correlation1} presents the correlations between various anomalous coupling paramters like $\bar c_{HW} - \bar c_{HB}$, $\bar c_{W} - \bar c_{HB}$, and $\bar c_{\gamma} - \bar c_{HB}$, where the yellow and gray bands show the present limits derived from the LHC results on the associated production of the Higgs boson with the $W$ boson \cite{Ellis:2014dva}. In the absence of any other parameter, the allowed region in the $\bar c_{HW} - \bar c_{HB}$ plane is restricted to a narrow ellipse (red). This ellipse is not affected much by the presence of $\bar c_W$ if it is positive (green ellipse). On the other hand, if $\bar c_W$ is negative, within the present bounds, it can significantly affect the allowed region (blue ellipse) in the $\bar c_{HW} - \bar c_{HB}$ plane. Similarly, the allowed region in the $\bar c_{W} - \bar c_{HB}$ and $\bar c_{\gamma} - \bar c_{HB}$ planes are illustrating the the presence of $\bar c_{HW}$ and $\bar c_{W}=-\bar c_B$ respectivelly.  

Presently, we would like to be content with the analysis at the production level, considering the limited scope of this work. However, the detector effects may lead to reduction in the sensitivities of the observables studied. It is thus needed to perform a full detector level simulation to estimate the realistic efficiencies of these observables. We have left this as a future work. As mentioned earlier, we shall focus on an ILC running at a center-of-mass energy of 500 GeV for our study. In order to understand  the interplay of various CP-conserving couplings, we consider $\bar c_W$, $\bar c_{HW}$, $\bar c_{HB}$ and $\bar c_{\gamma}$ parameters related to anomalous couplings. 

\begin{figure}[H] \centering
\begin{tabular}{c c}
\hspace{1mm}
\includegraphics[angle=0,width=80mm]{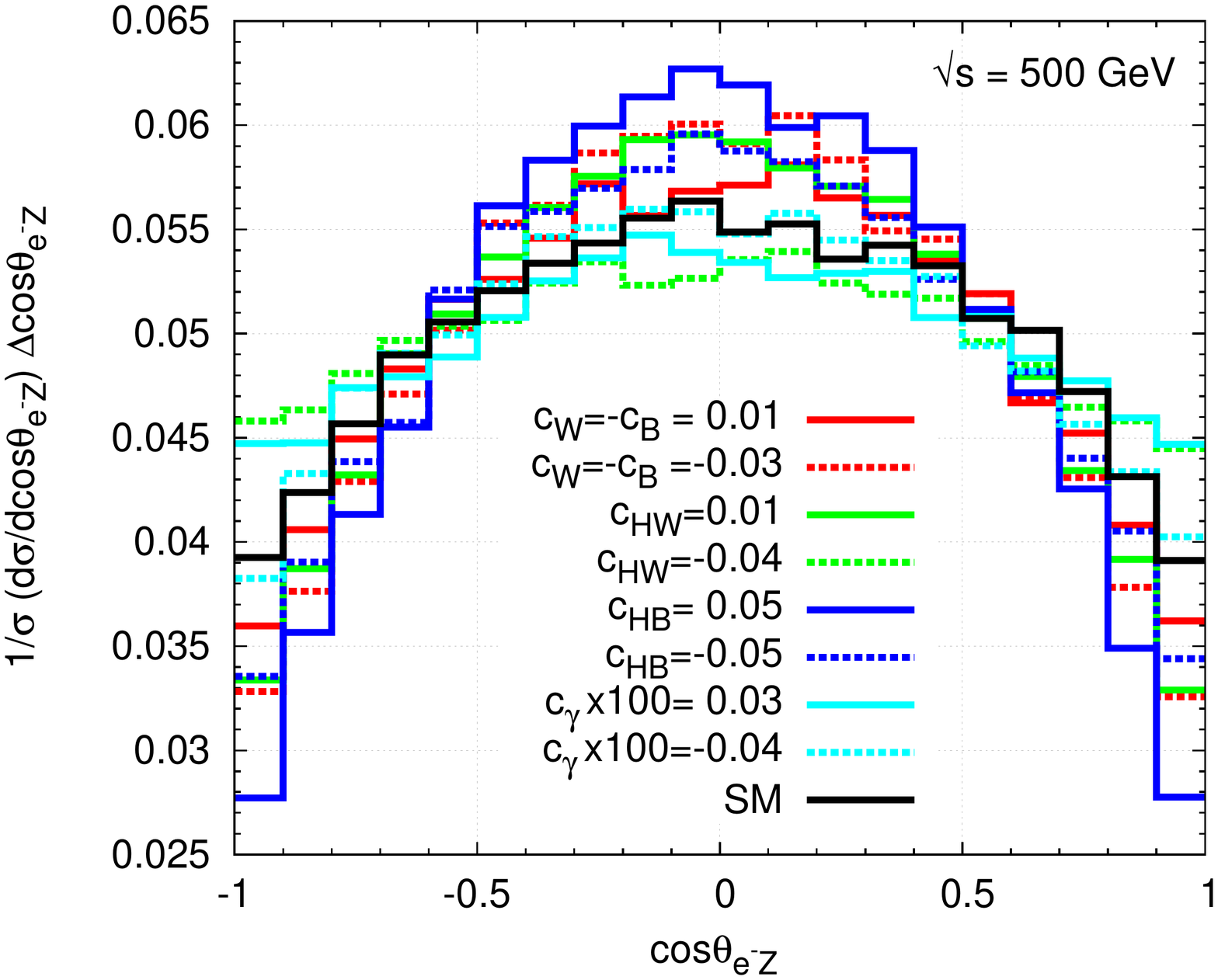} &
\hspace{-10mm}
\includegraphics[angle=0,width=80mm]{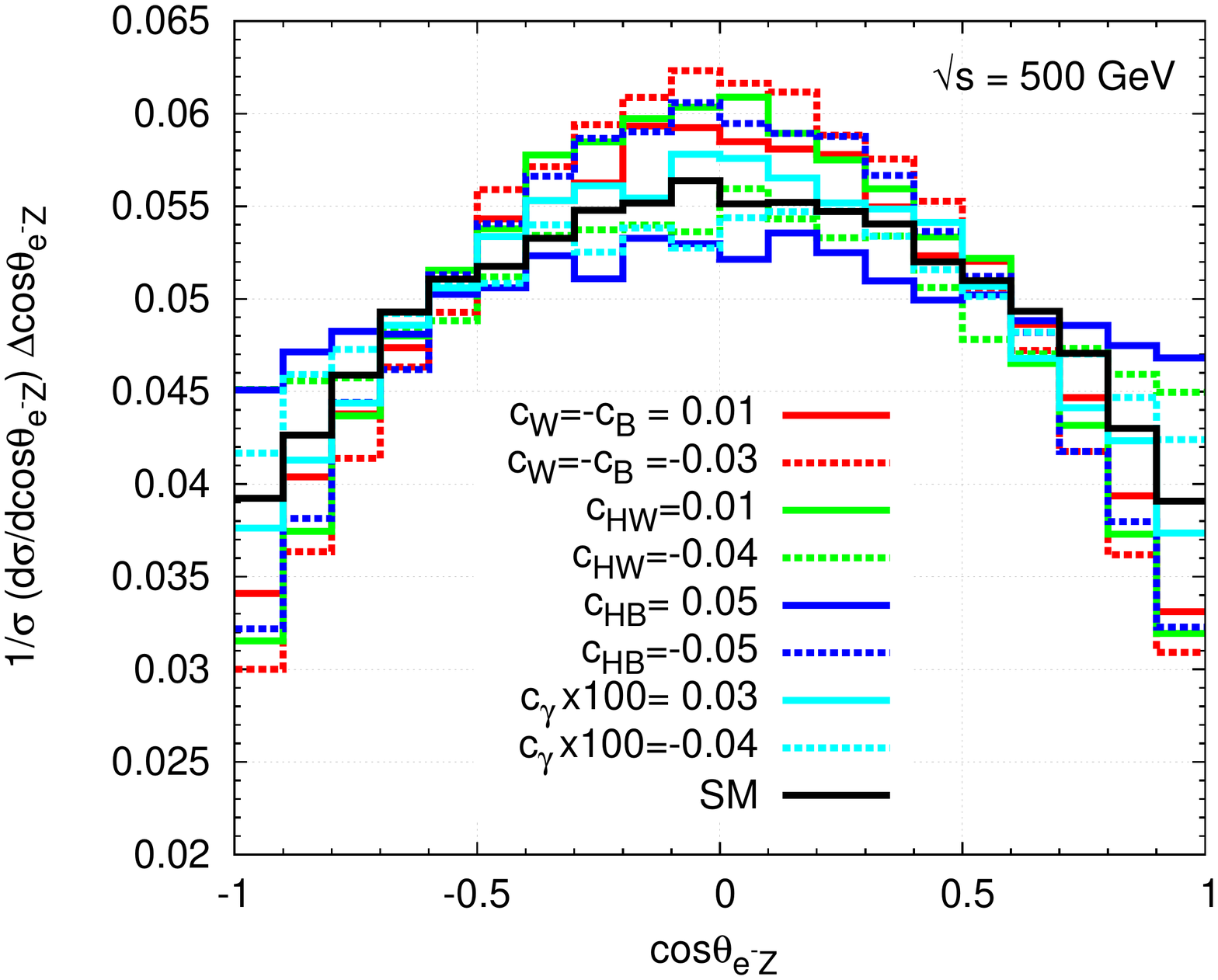} \\
\end{tabular}
\vspace{-12mm}
\caption{Distribution of $\cos\theta_Z$ for different anomalous couplings with unpolarized (left) and polarized with $P_{e^-}=-80\%,~~P_{e^+}=+30\%$ beams (right). A center-of-mass energy of $500$ GeV is assumed.}
\label{fig:angZ}
\end{figure}

\begin{figure}[H] \centering
\begin{tabular}{c c}
\hspace{1mm}
\includegraphics[angle=0,width=80mm]{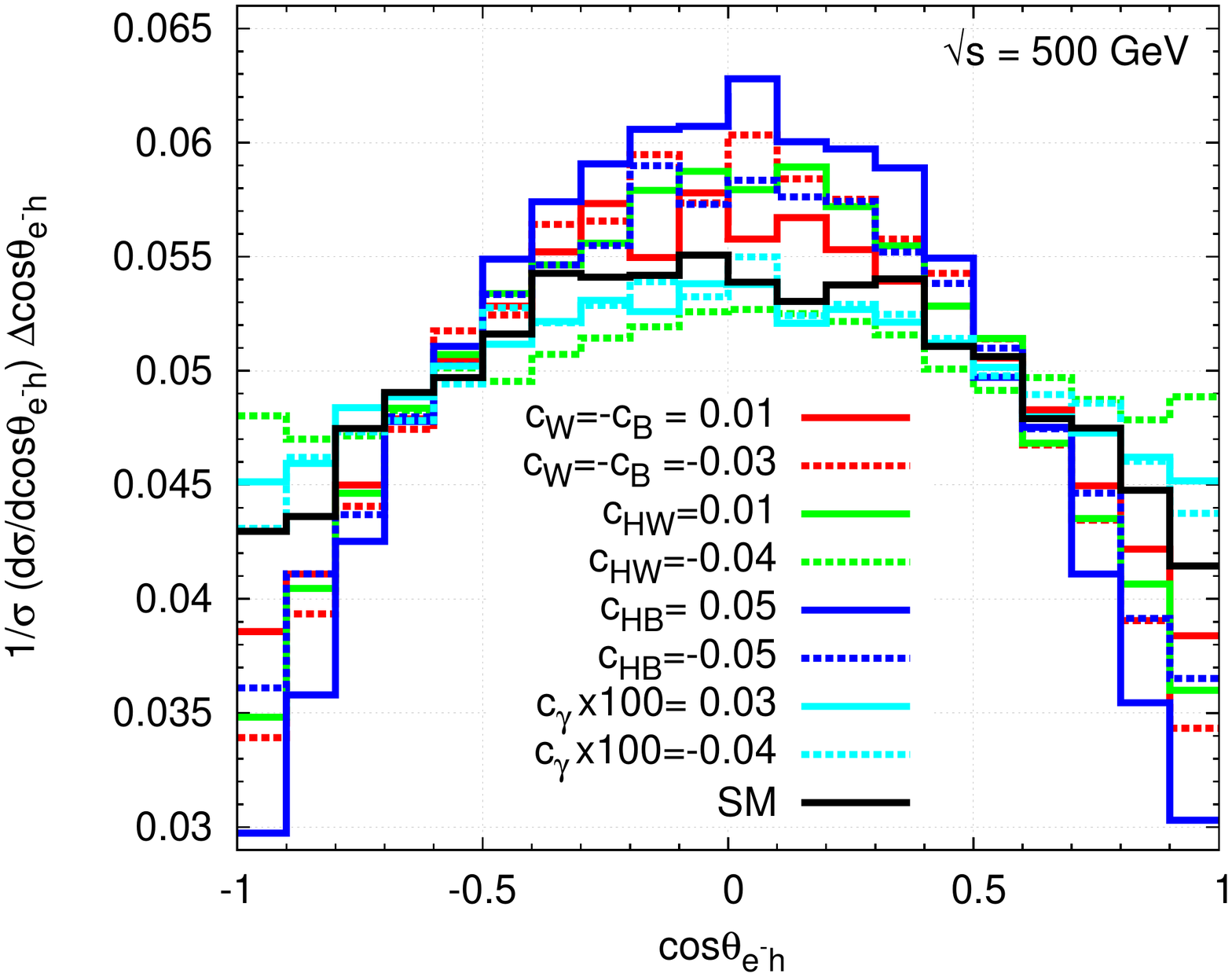} &
\hspace{-10mm}
\includegraphics[angle=0,width=80mm]{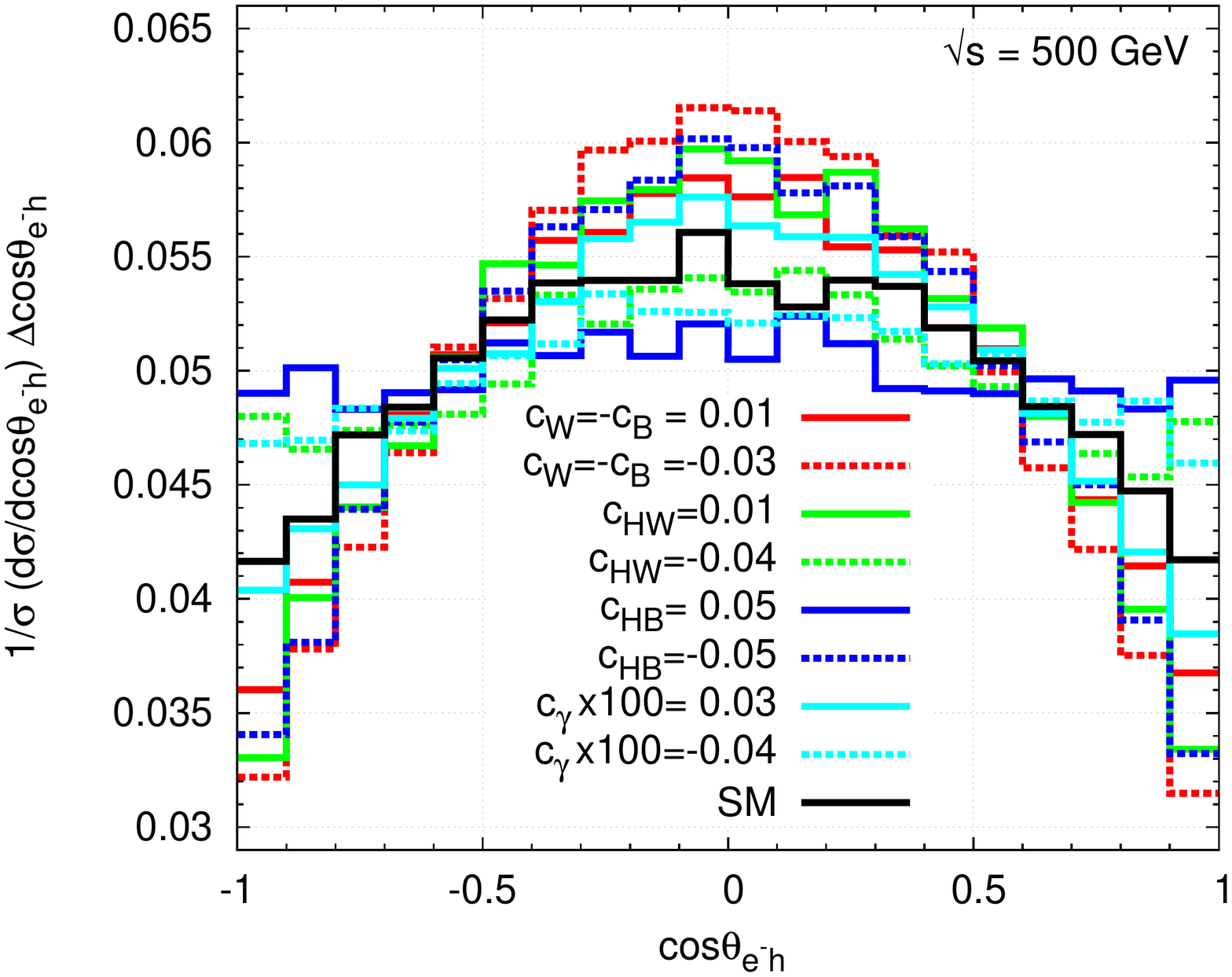} \\
\end{tabular}
\vspace{-12mm}
\caption{Distribution of $\cos\theta_h$ for different anomalous couplings with unpolarized (left) and polarized with $P_{e^-}=-80\%,~~P_{e^+}=+30\%$ beams (right). A center-of-mass energy of $500$ GeV is assumed.}
\label{fig:angH}
\end{figure}

\begin{figure}[H] \centering
\begin{tabular}{c c}
\hspace{1mm}
\includegraphics[angle=0,width=80mm]{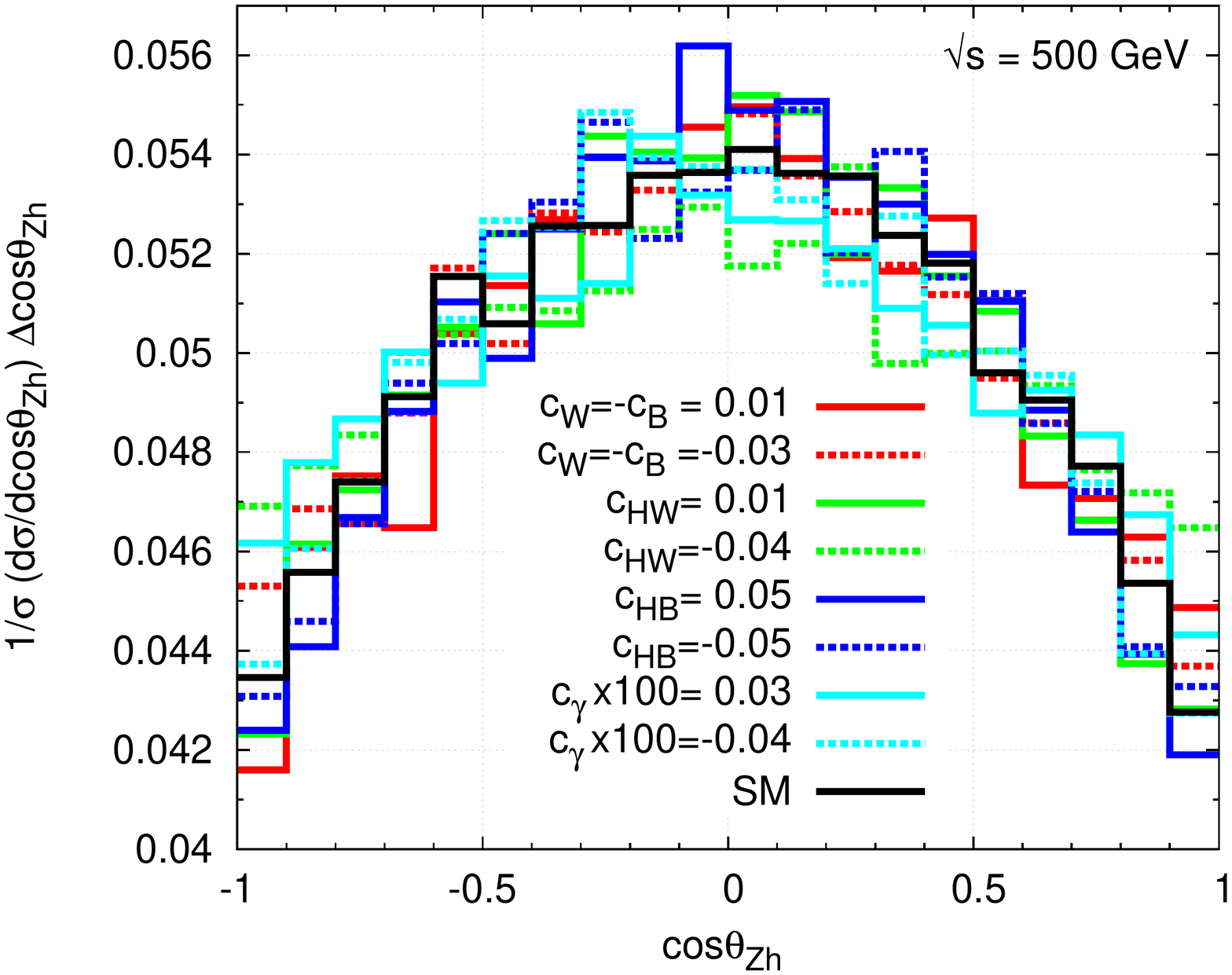} &
\vspace{-8mm}\hspace{-10mm}
\includegraphics[angle=0,width=80mm]{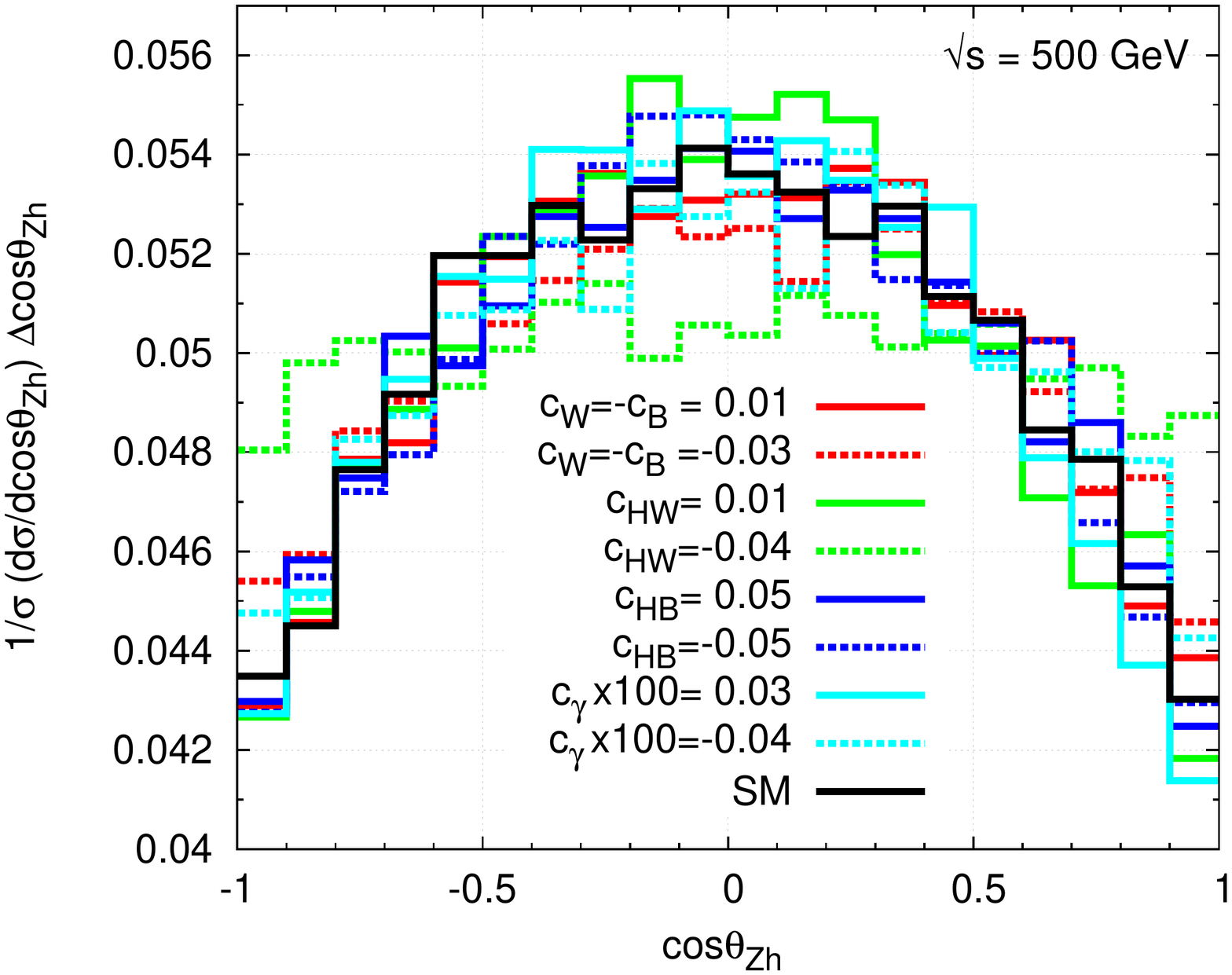} \\
\hspace{1mm}
\includegraphics[angle=0,width=80mm]{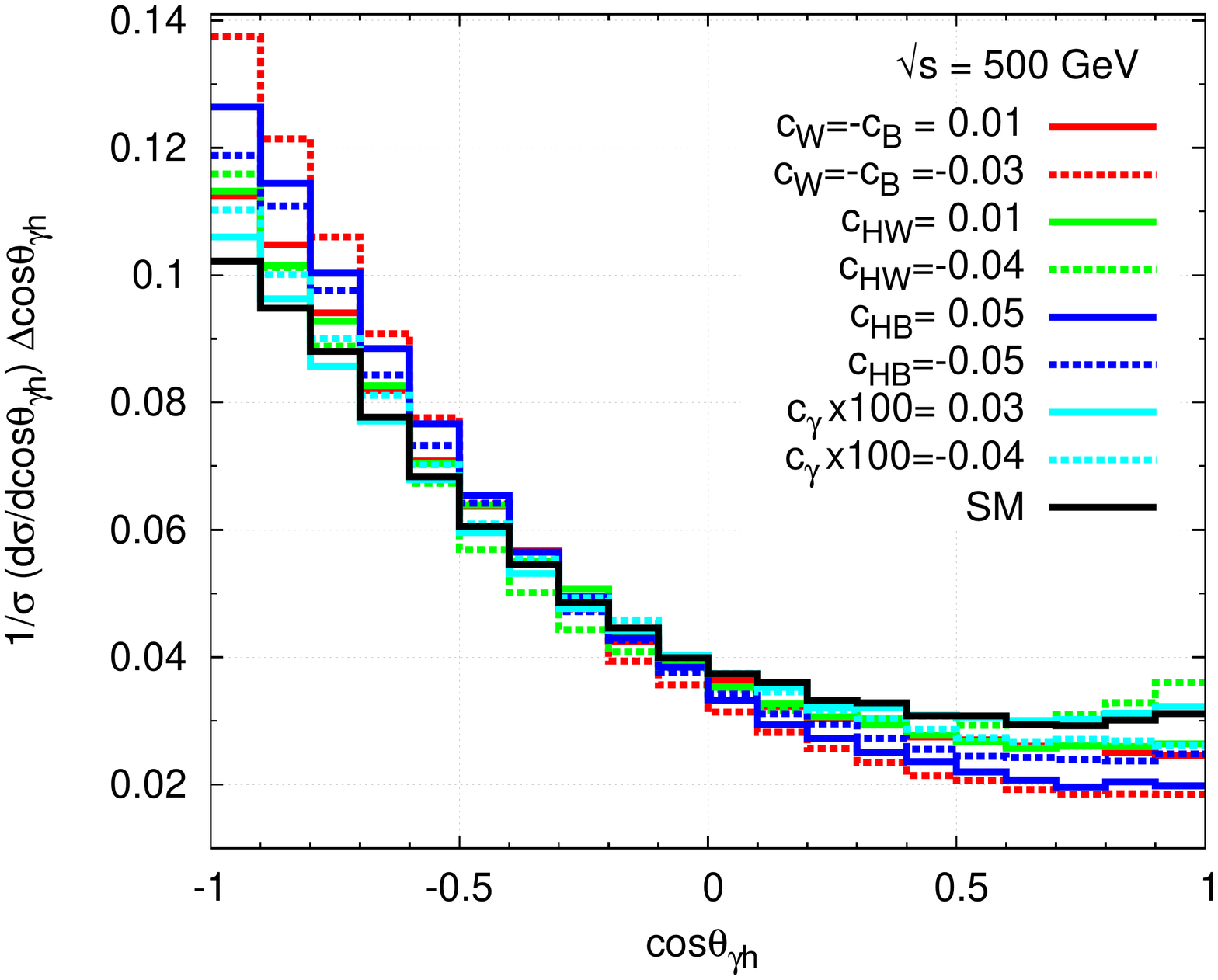} &
\vspace{-8mm}\hspace{-10mm}
\includegraphics[angle=0,width=80mm]{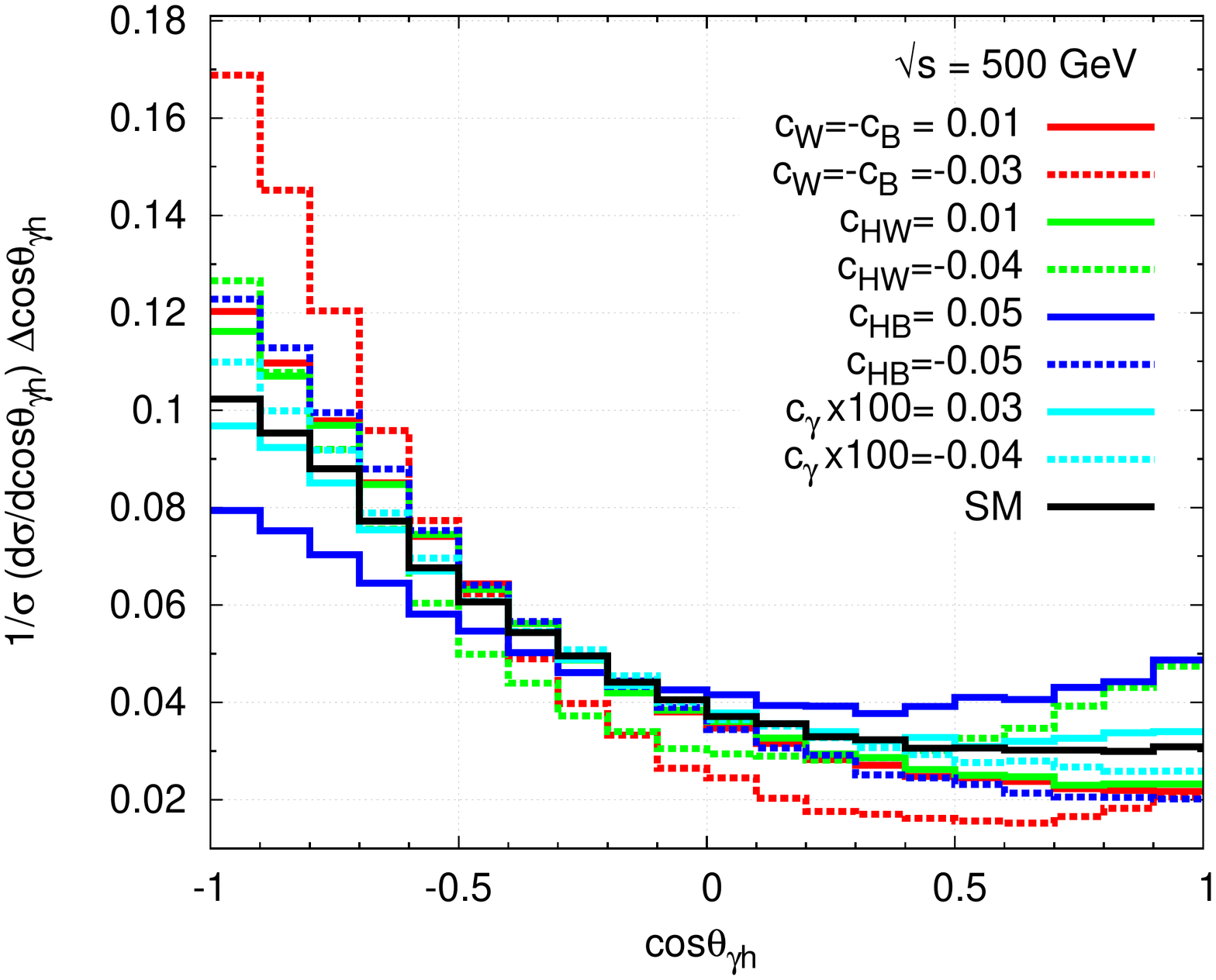} \\
\hspace{1mm}
\includegraphics[angle=0,width=80mm]{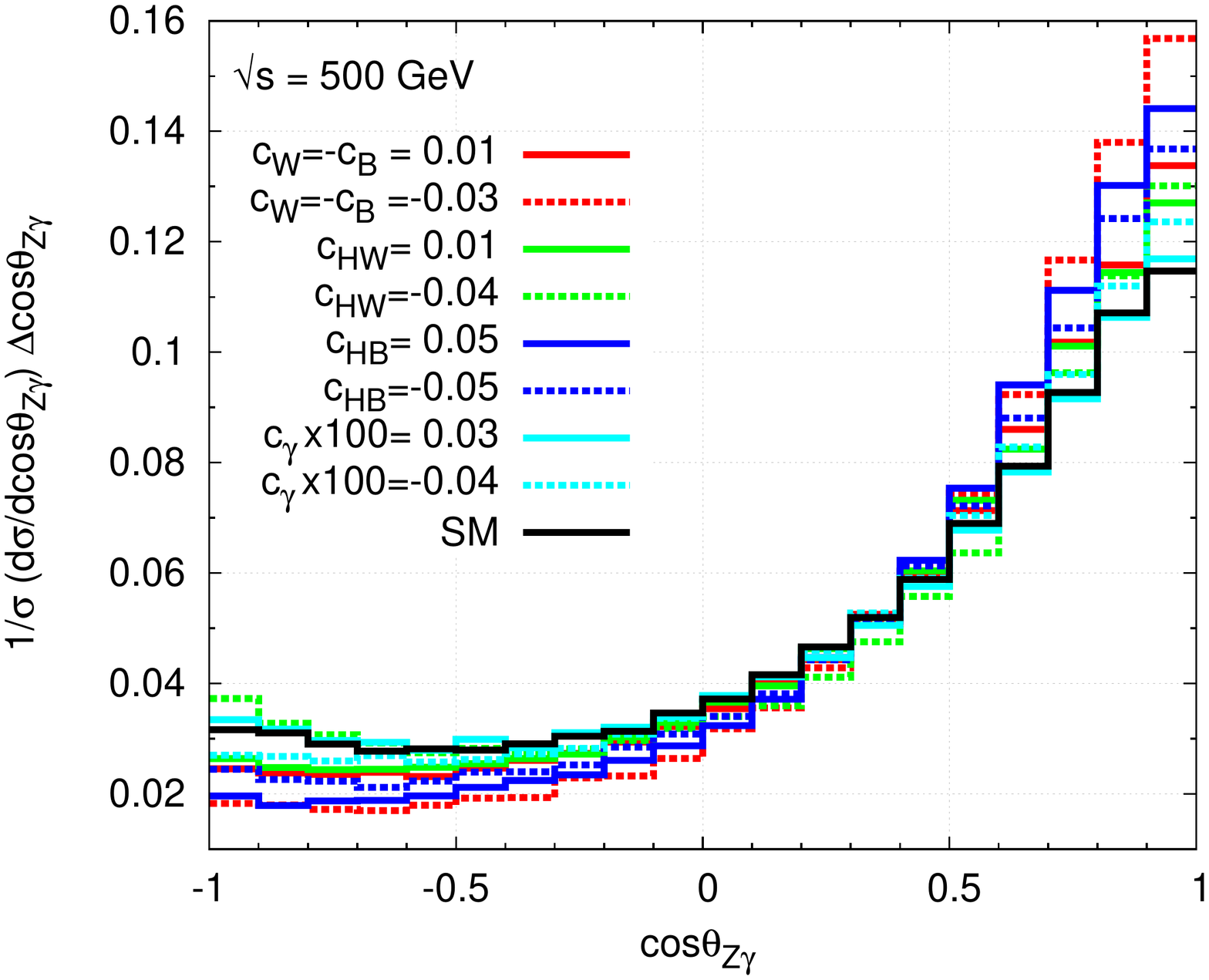}  & 
\hspace{-10mm}
\includegraphics[angle=0,width=80mm]{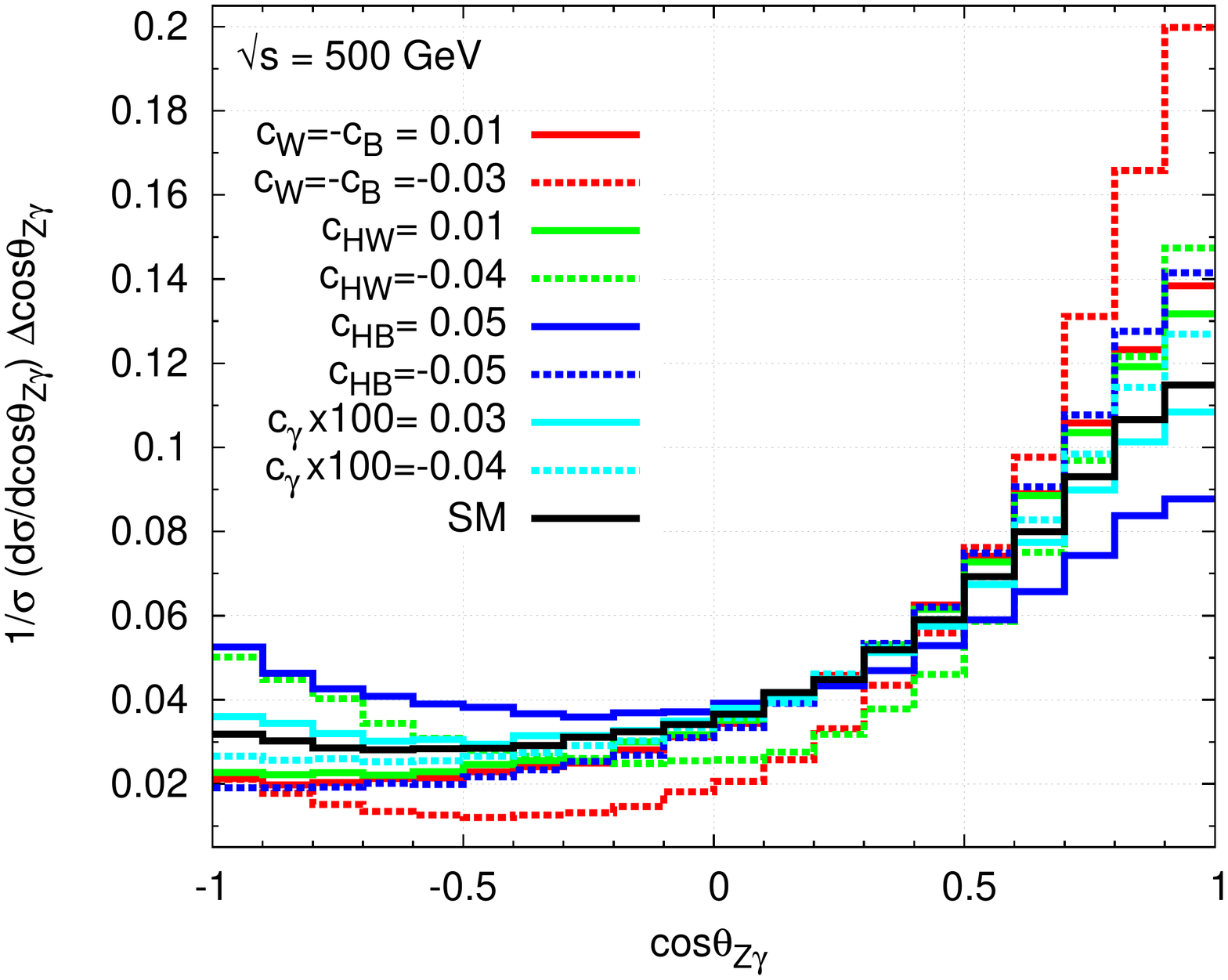} \\
\end{tabular}
\vspace{-10mm}
\caption{The normalized distributions of $\cos\theta_{Zh}$ (top), $\cos\theta_{\gamma h}$ (middle) and $\cos\theta_{Z \gamma}$ for different anomalous couplings with unpolarized (left) and polarized with $P_{e^-}=-80\%,~~P_{e^+}=+30\%$ beams (right). A center-of-mass energy of $500$ GeV is assumed.}
\label{fig:angZH_AH_ZA}
\end{figure}

\begin{figure}[H] \centering
\begin{tabular}{c c}
\hspace{1mm}
\includegraphics[angle=0,width=80mm]{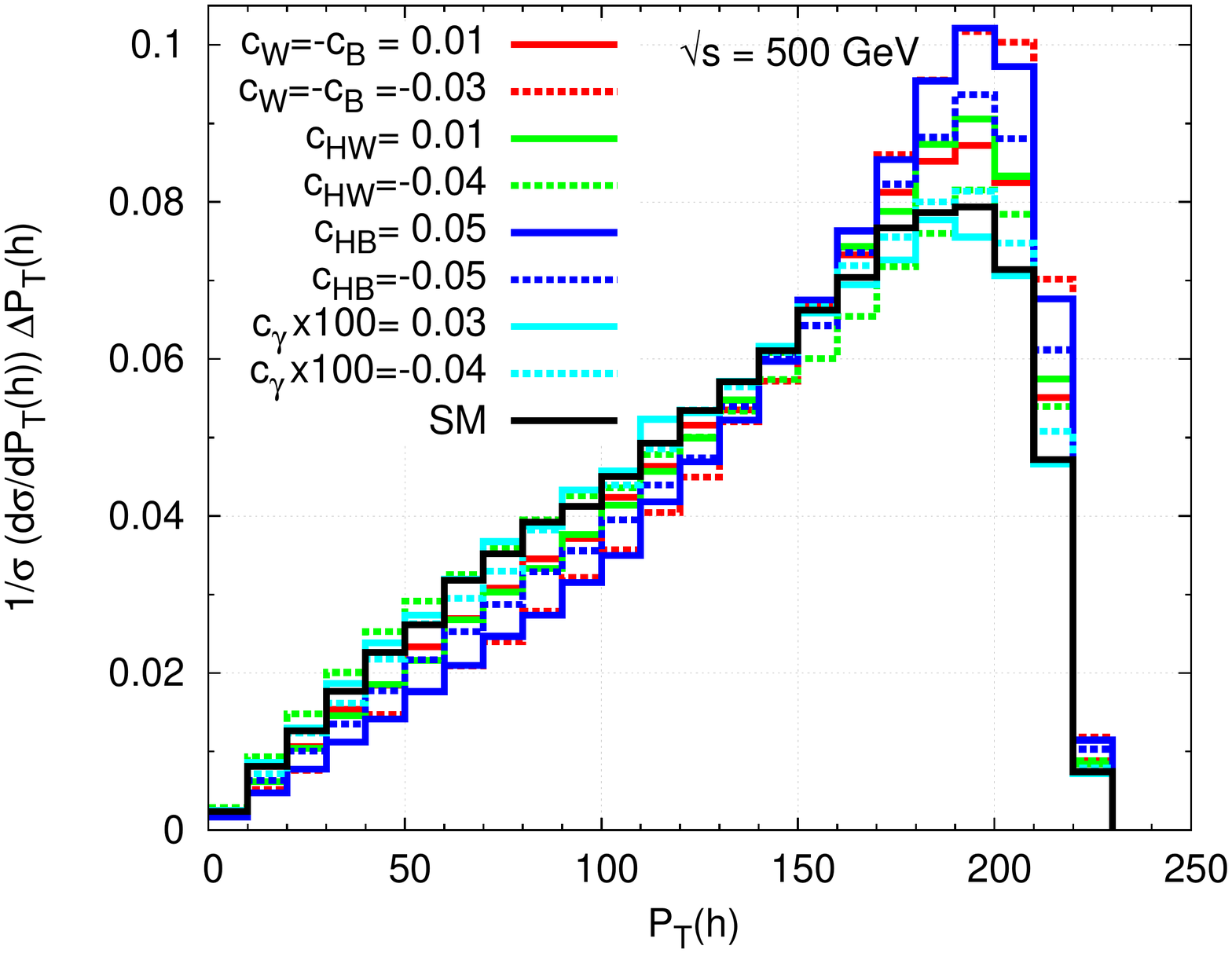} &
\vspace{-8mm}\hspace{-10mm}
\includegraphics[angle=0,width=80mm]{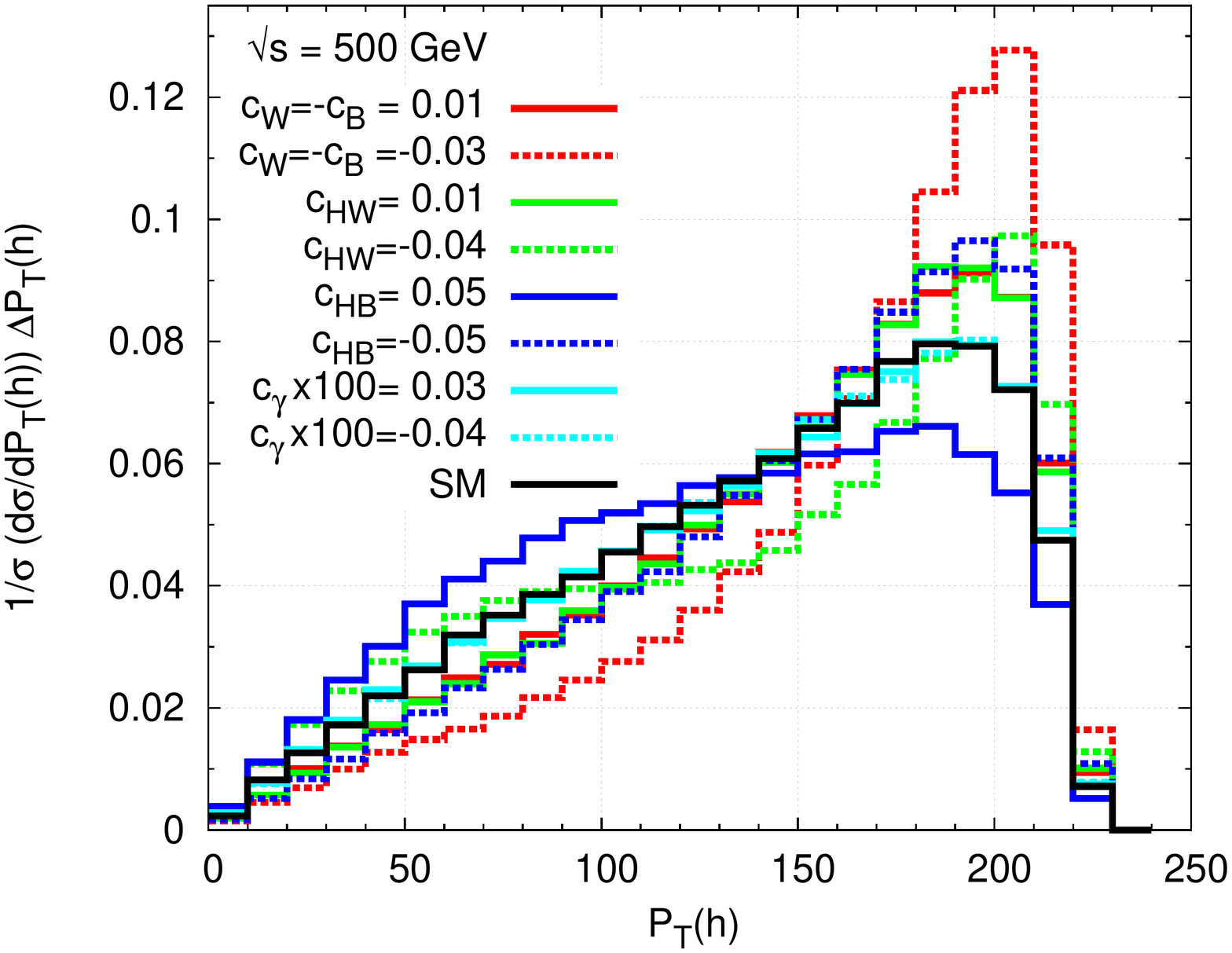} \\
\hspace{1mm}
\includegraphics[angle=0,width=80mm]{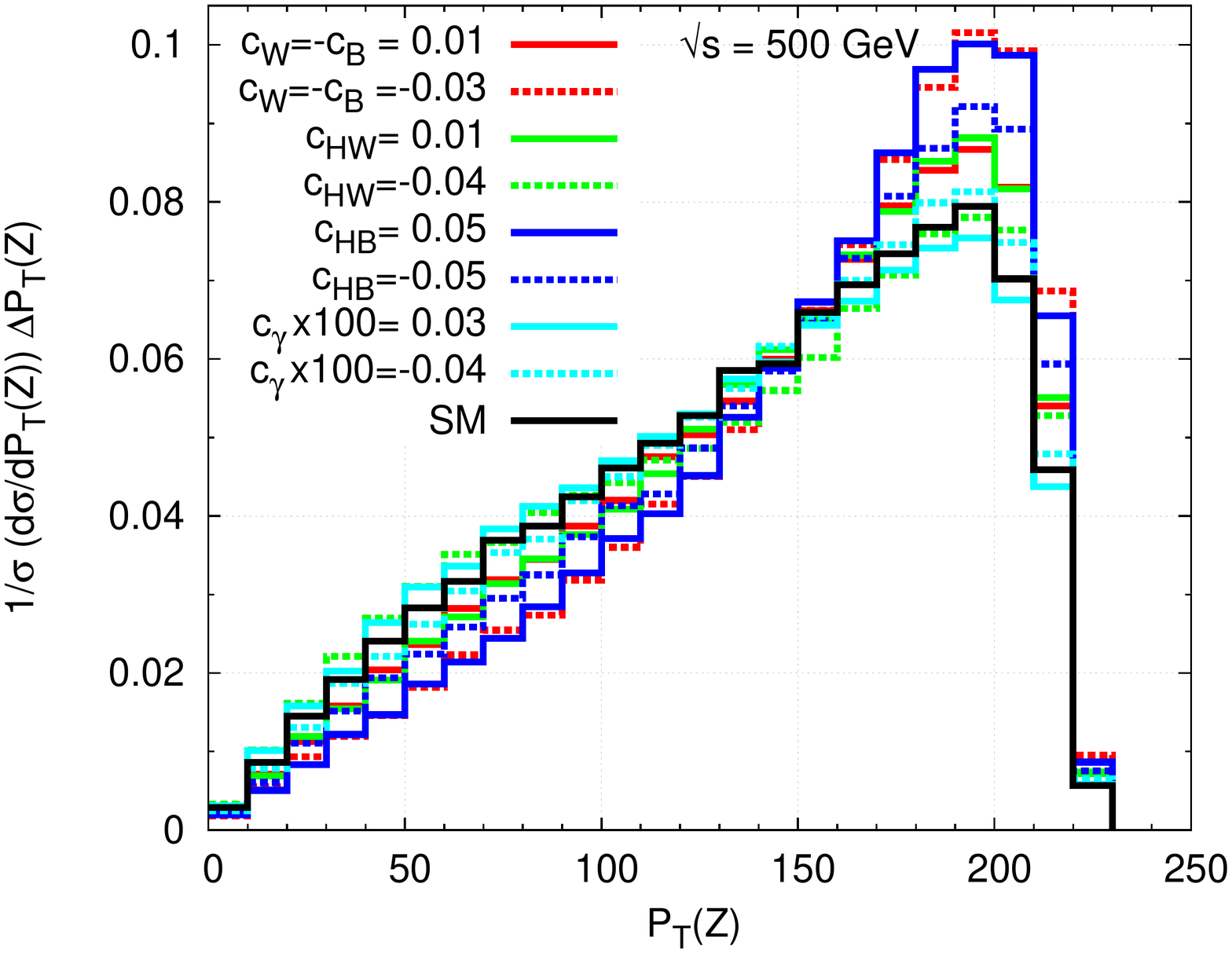} &
\vspace{-8mm}\hspace{-10mm}
\includegraphics[angle=0,width=80mm]{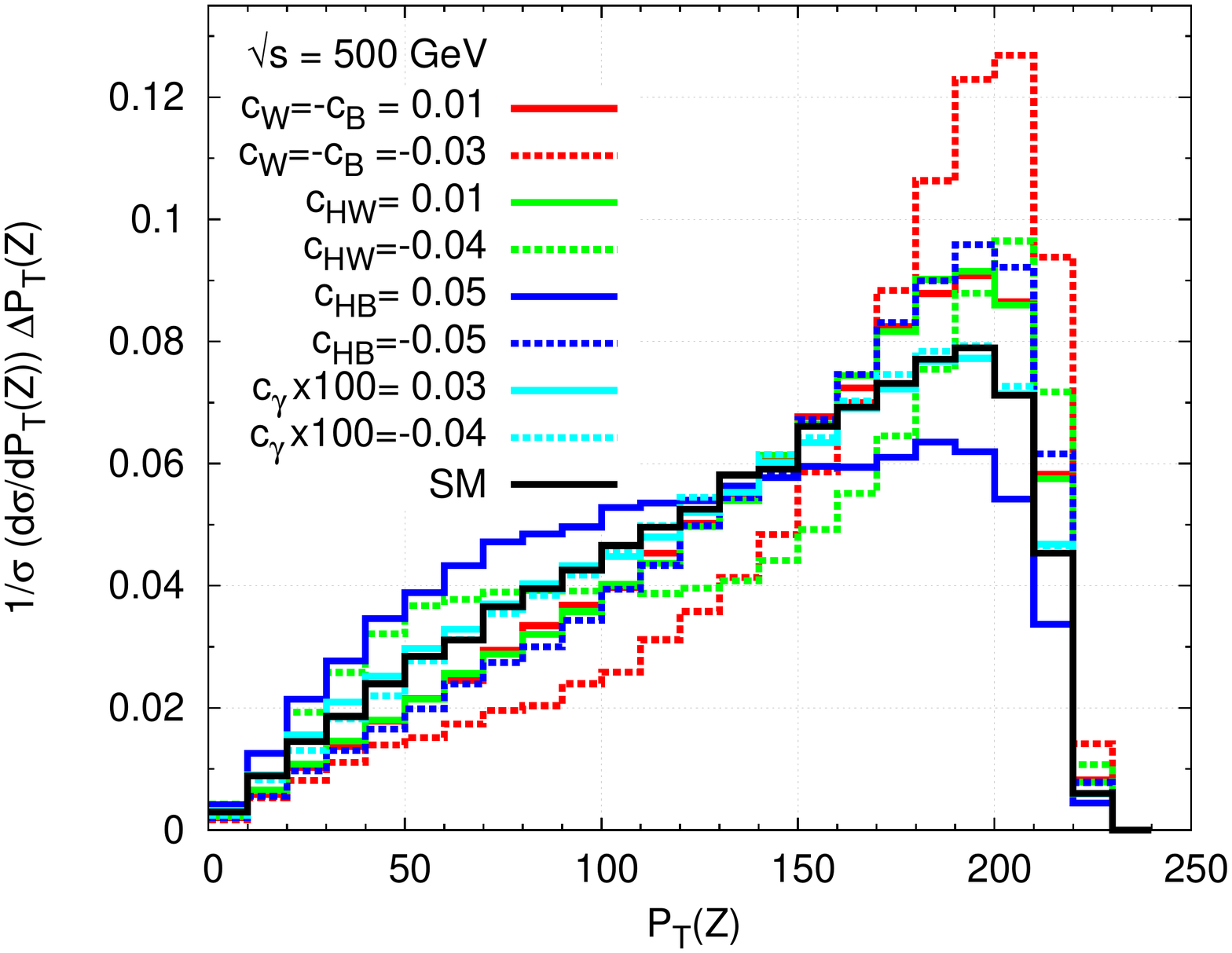} \\
\hspace{1mm}
\includegraphics[angle=0,width=80mm]{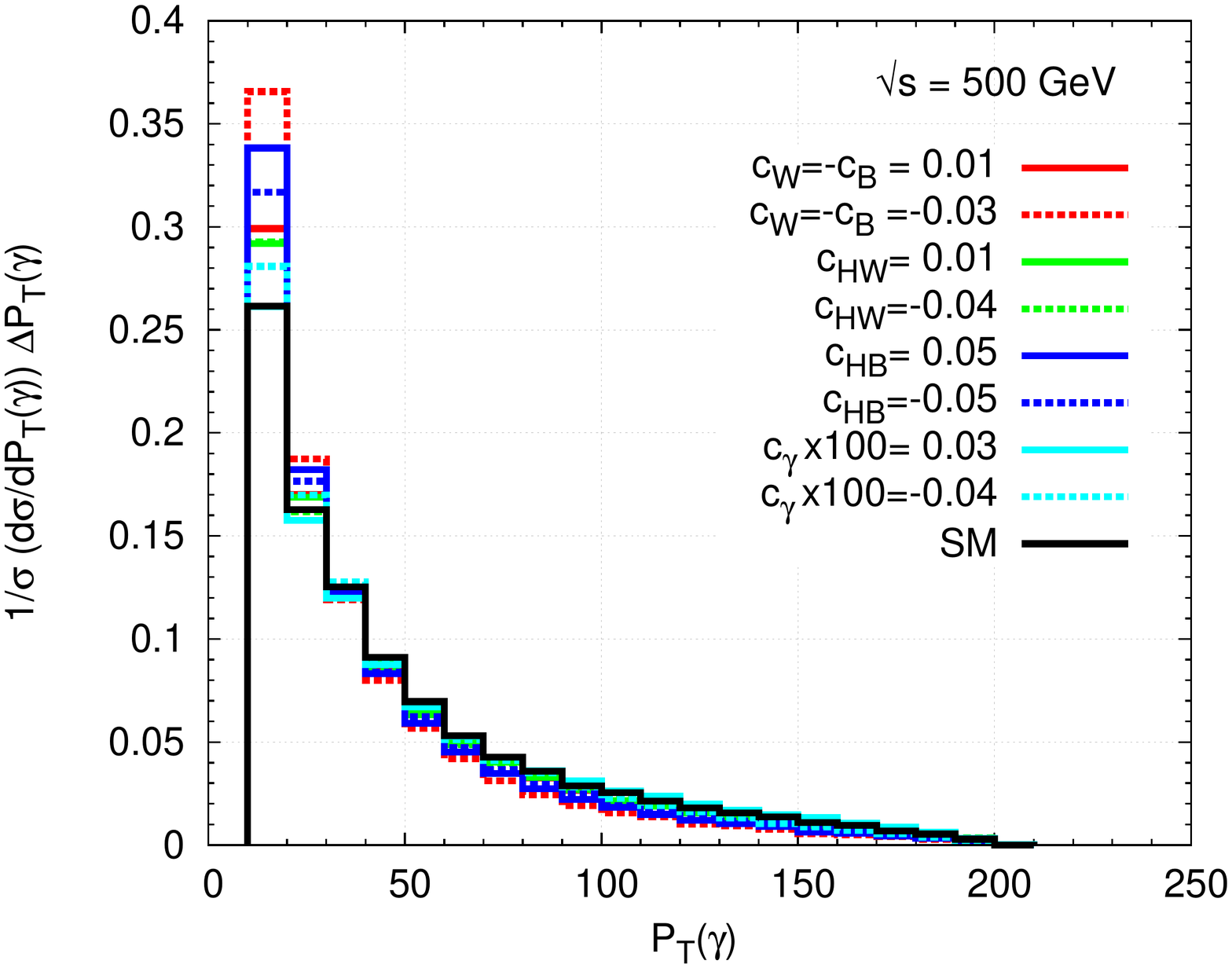} &
\hspace{-10mm}
\includegraphics[angle=0,width=80mm]{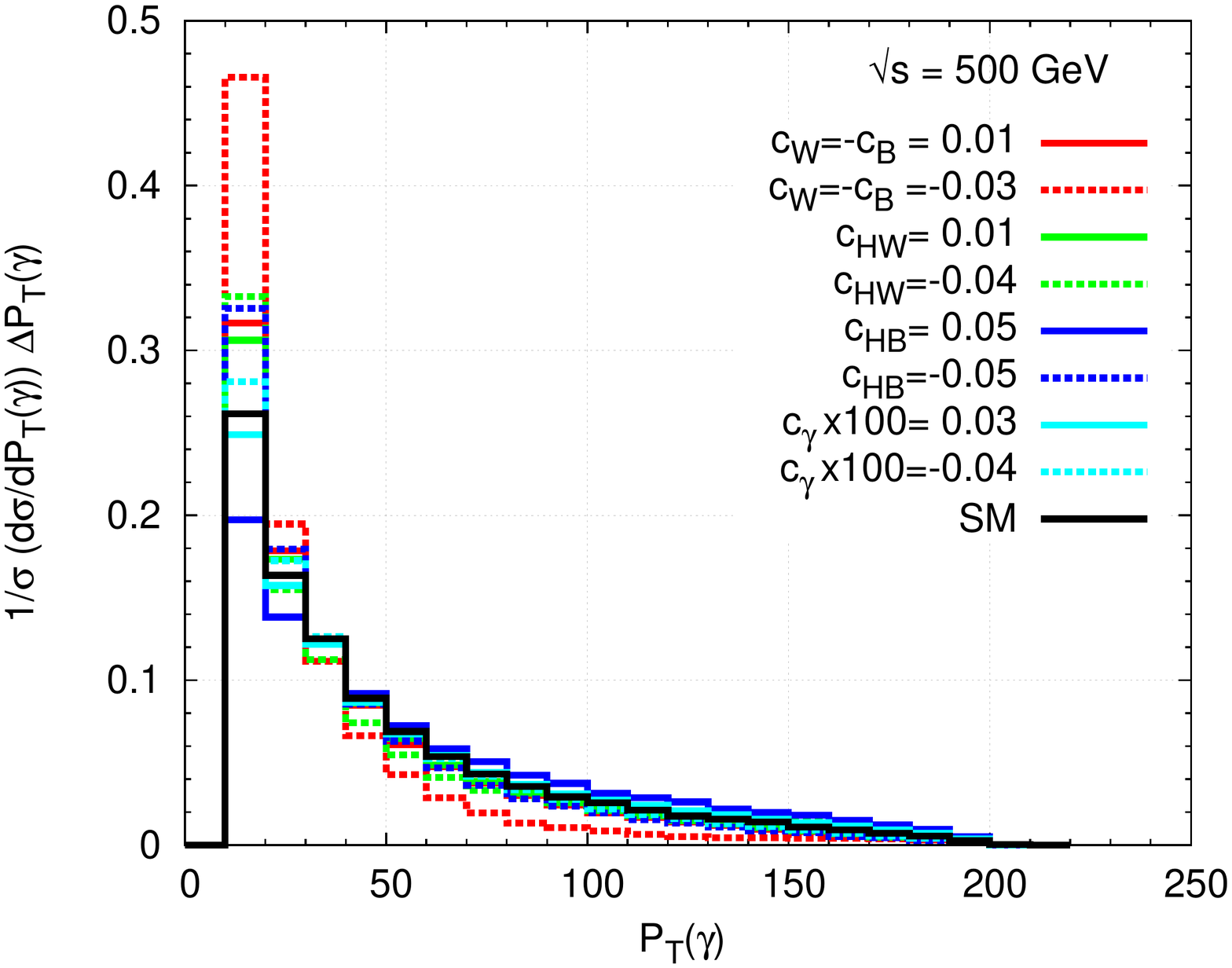} \\
\end{tabular}
\vspace{-10mm}
\caption{The normalized $P_T$ distributions of the Higgs boson (top), Z-boson (middle) and photon (bottom) for different anomalous couplings with unpolarized beams (left) and polarized with $P_{e^-}=-80\%,~~P_{e^+}=+30\%$ beams (right). A center-of-mass energy of $500$ GeV is assumed. A center-of-mass energy of $500$ GeV is assumed.}
\label{fig:Pt}
\end{figure}

\begin{figure}\centering
\begin{tabular}{c c}
\hspace{1mm}
\includegraphics[angle=0,width=80mm]{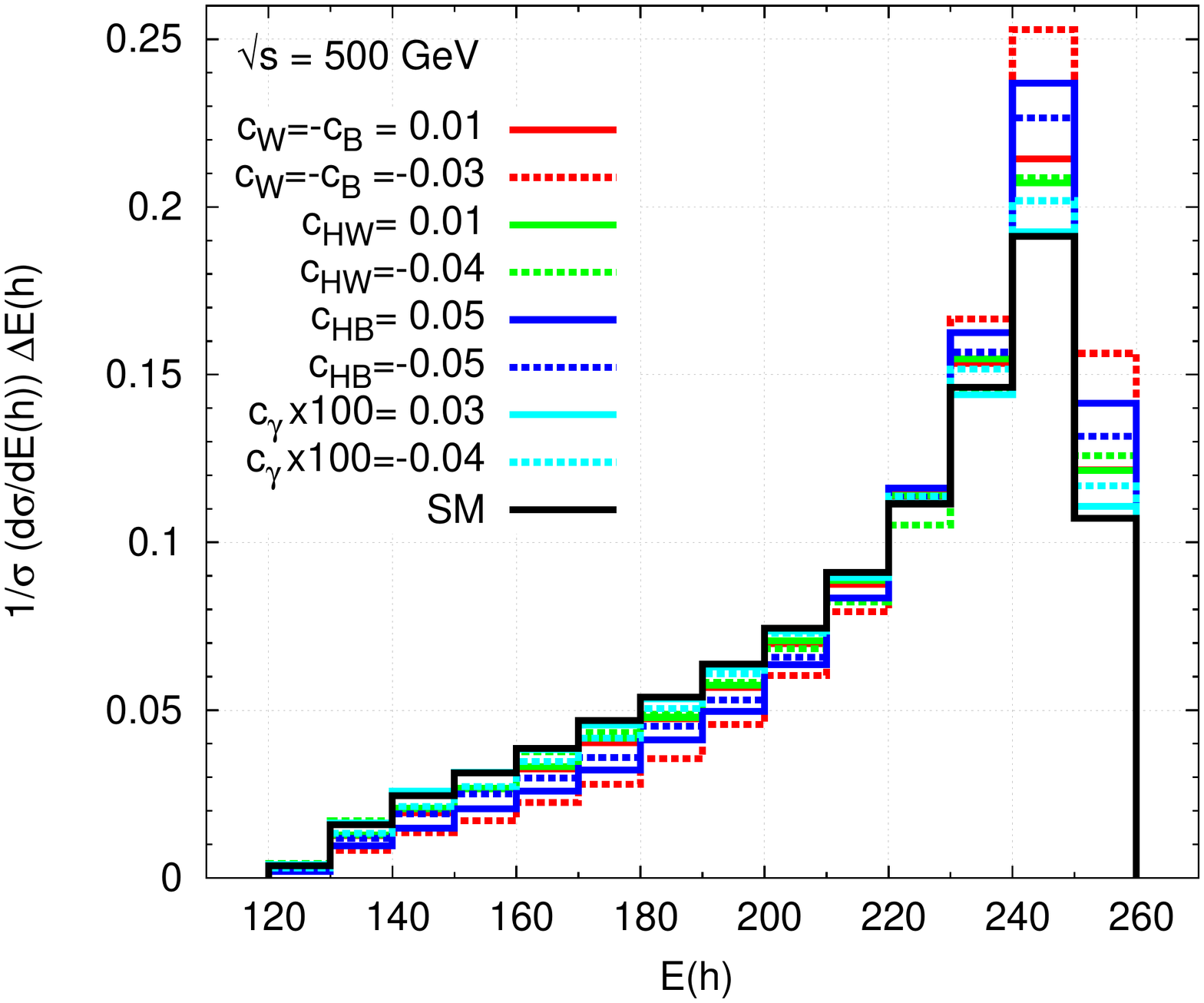} &
\vspace{-8mm}\hspace{-10mm}
\includegraphics[angle=0,width=80mm]{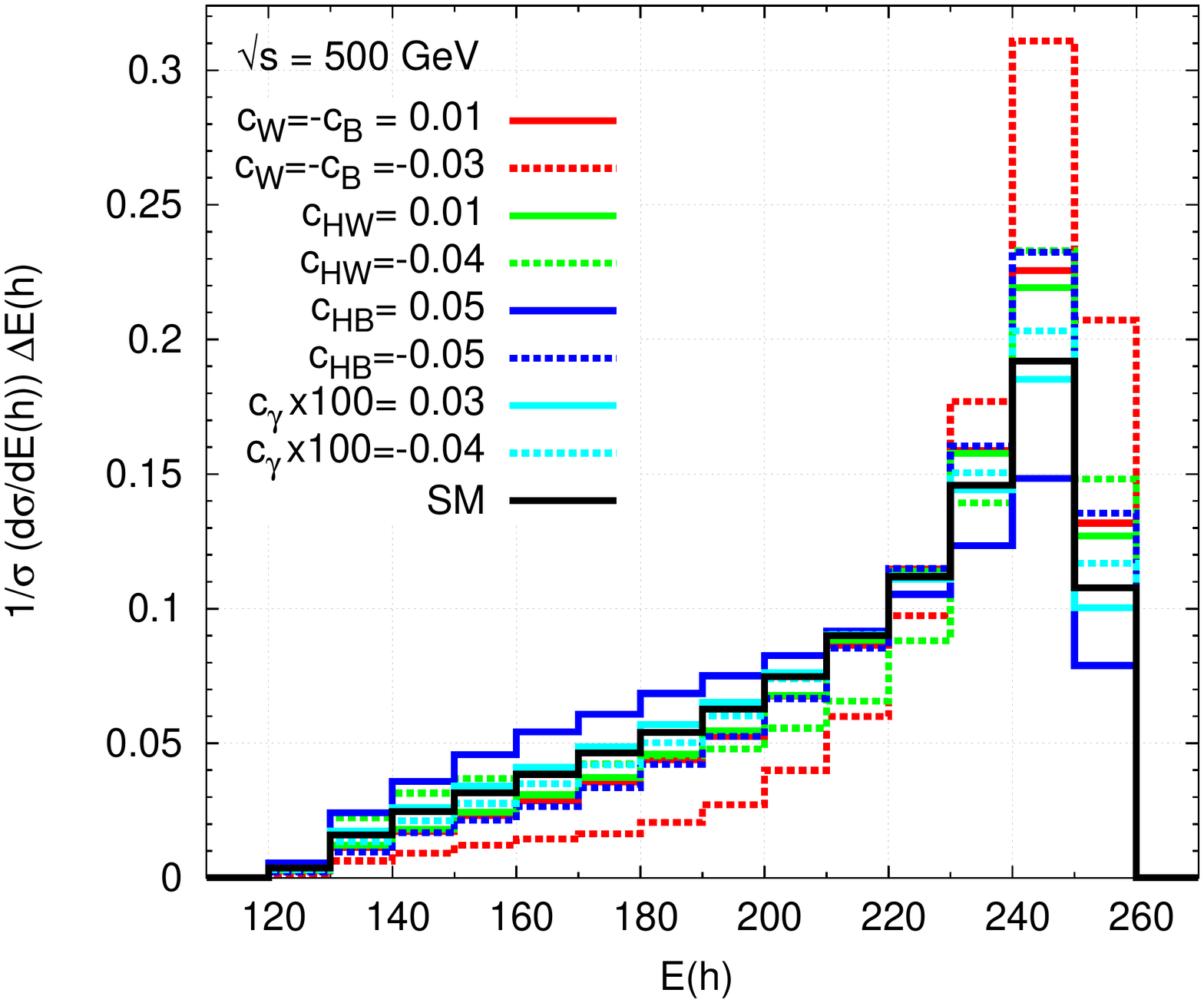} \\
\hspace{1mm}
\includegraphics[angle=0,width=80mm]{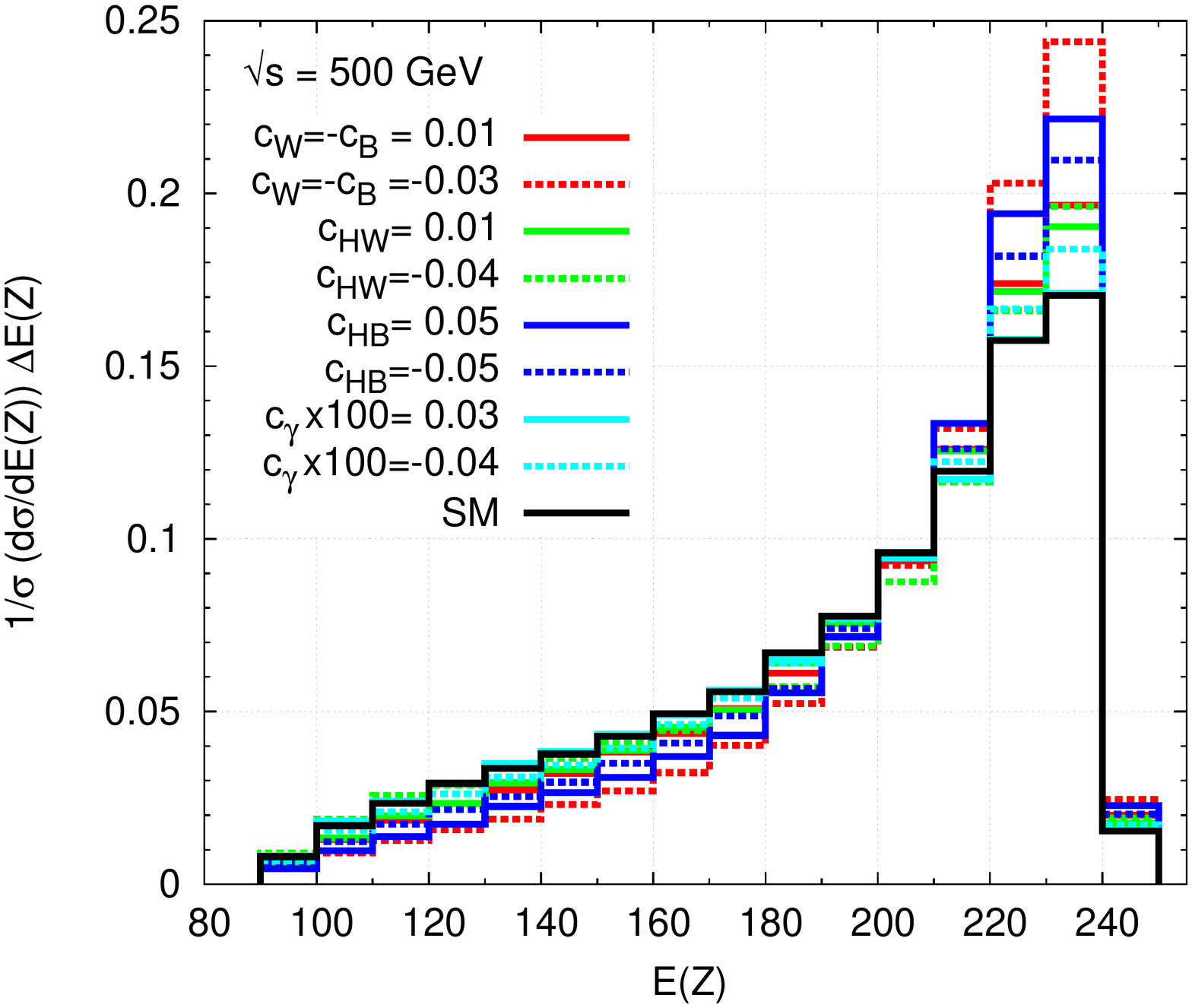} &
\vspace{-8mm}\hspace{-10mm}
\includegraphics[angle=0,width=80mm]{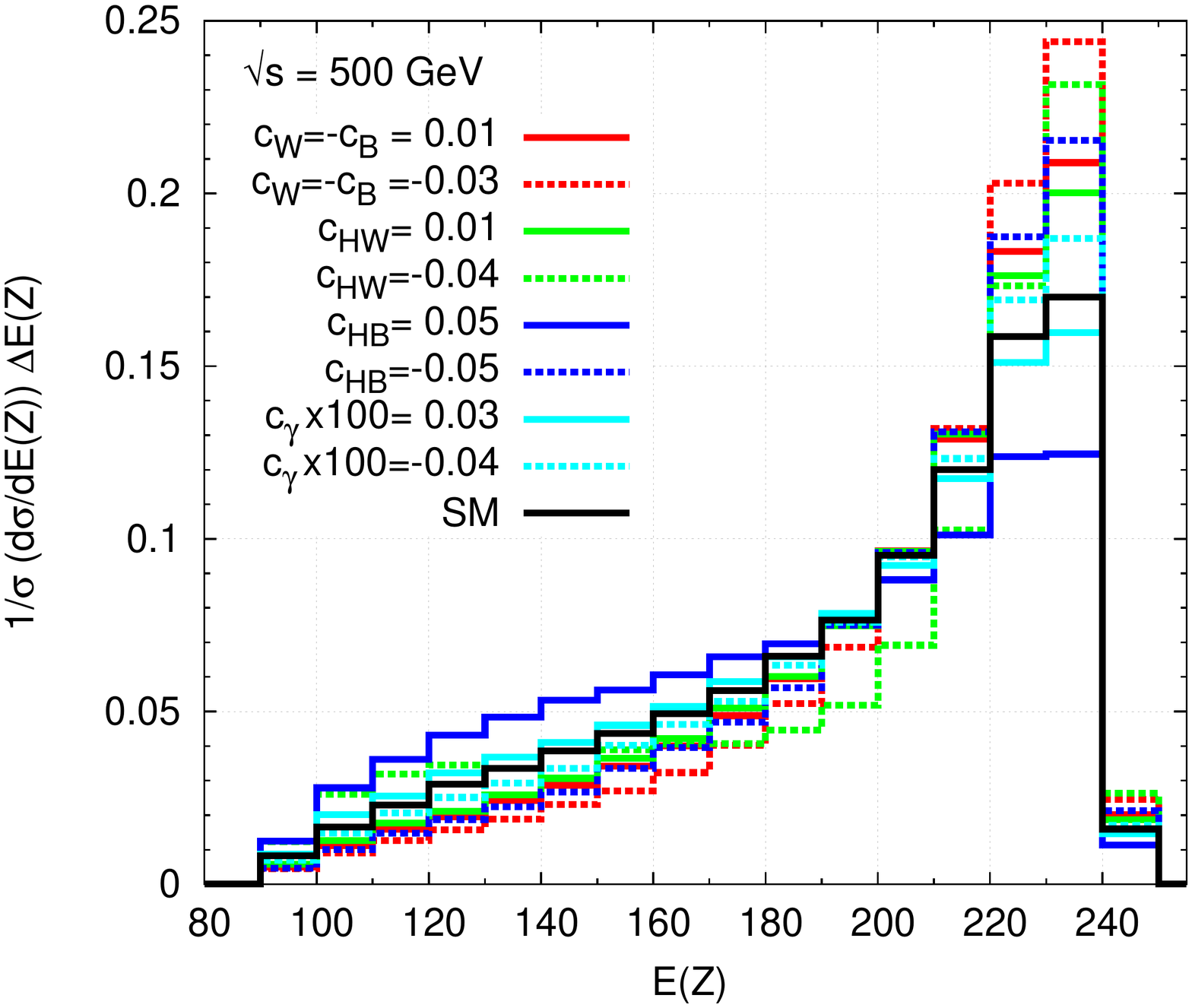} \\
\hspace{1mm}
\includegraphics[angle=0,width=80mm]{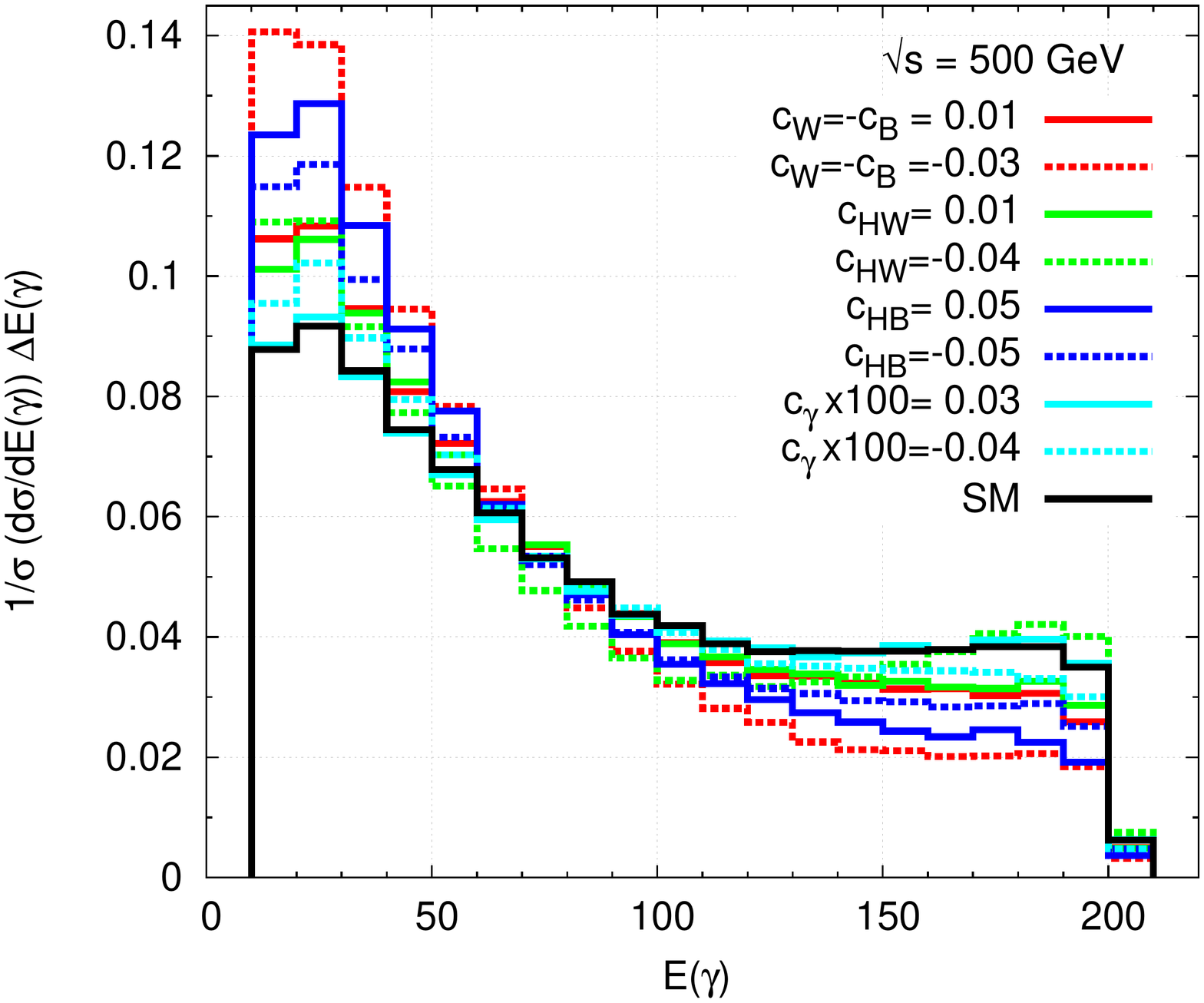} &
\hspace{-10mm}
\includegraphics[angle=0,width=80mm]{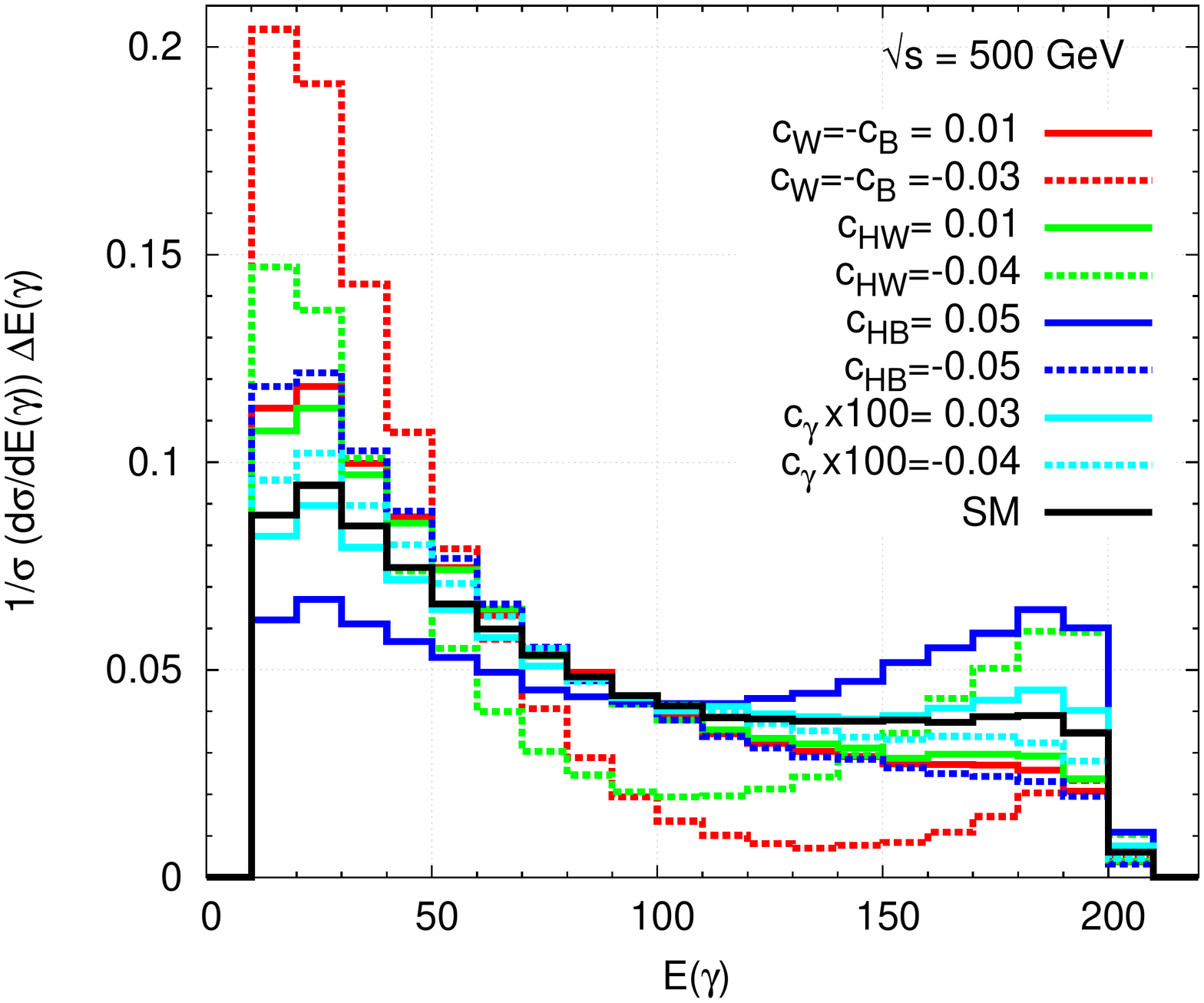} \\
\end{tabular}
\vspace{-10mm}
\caption{The normalized energy distributions of the Higgs boson (top), Z-boson (middle) and photon (bottom) for different anomalous couplings with unpolarized (left) and polarized with $P_{e^-}=-80\%,~~P_{e^+}=+30\%$ (right) beams. A center-of-mass energy of $500$ GeV is assumed.}
\label{fig:energy}
\end{figure}

\begin{figure}\centering
\begin{tabular}{c c}
\hspace{1mm}
\includegraphics[angle=0,width=80mm]{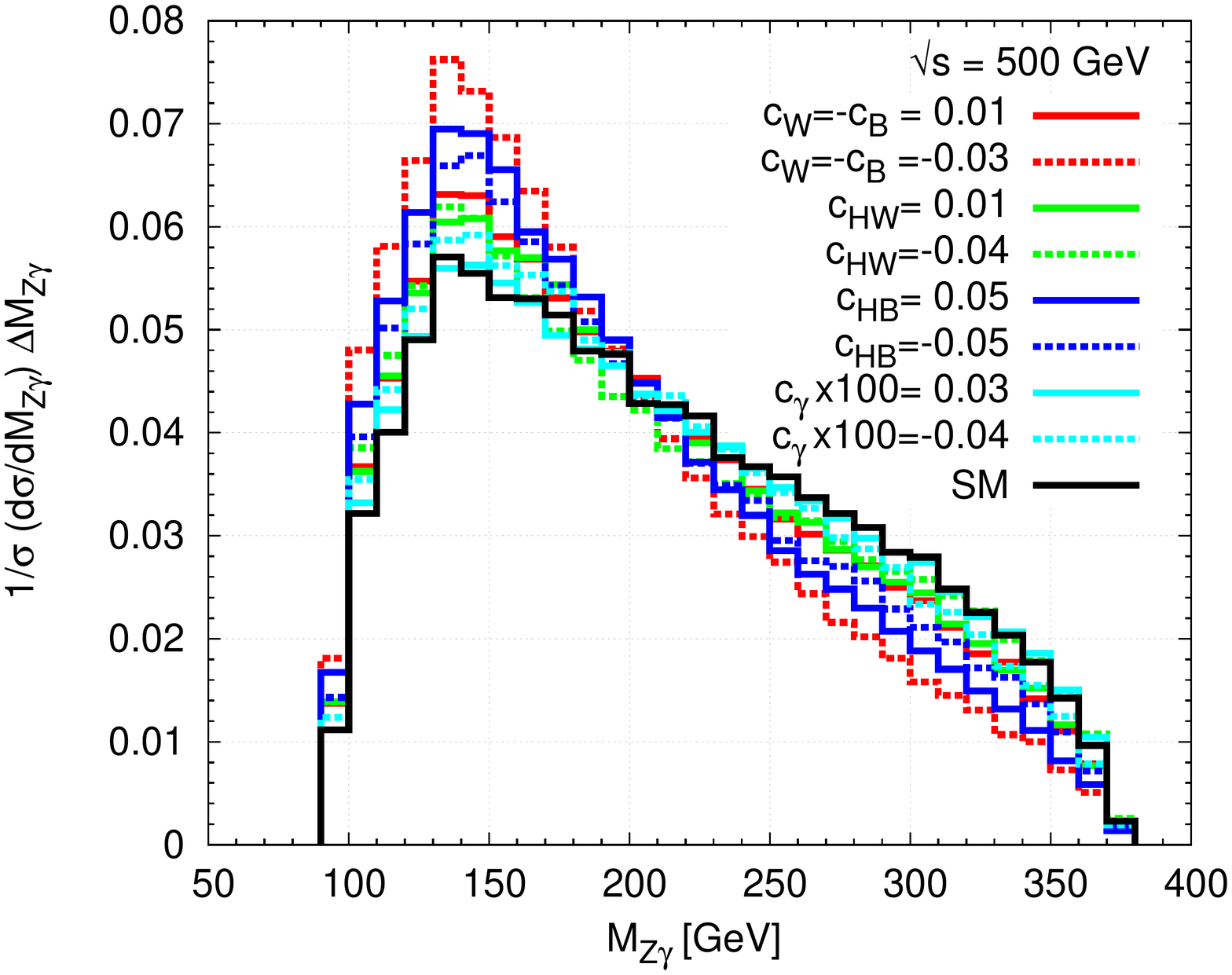} &
\vspace{-8mm}\hspace{-10mm}
\includegraphics[angle=0,width=80mm]{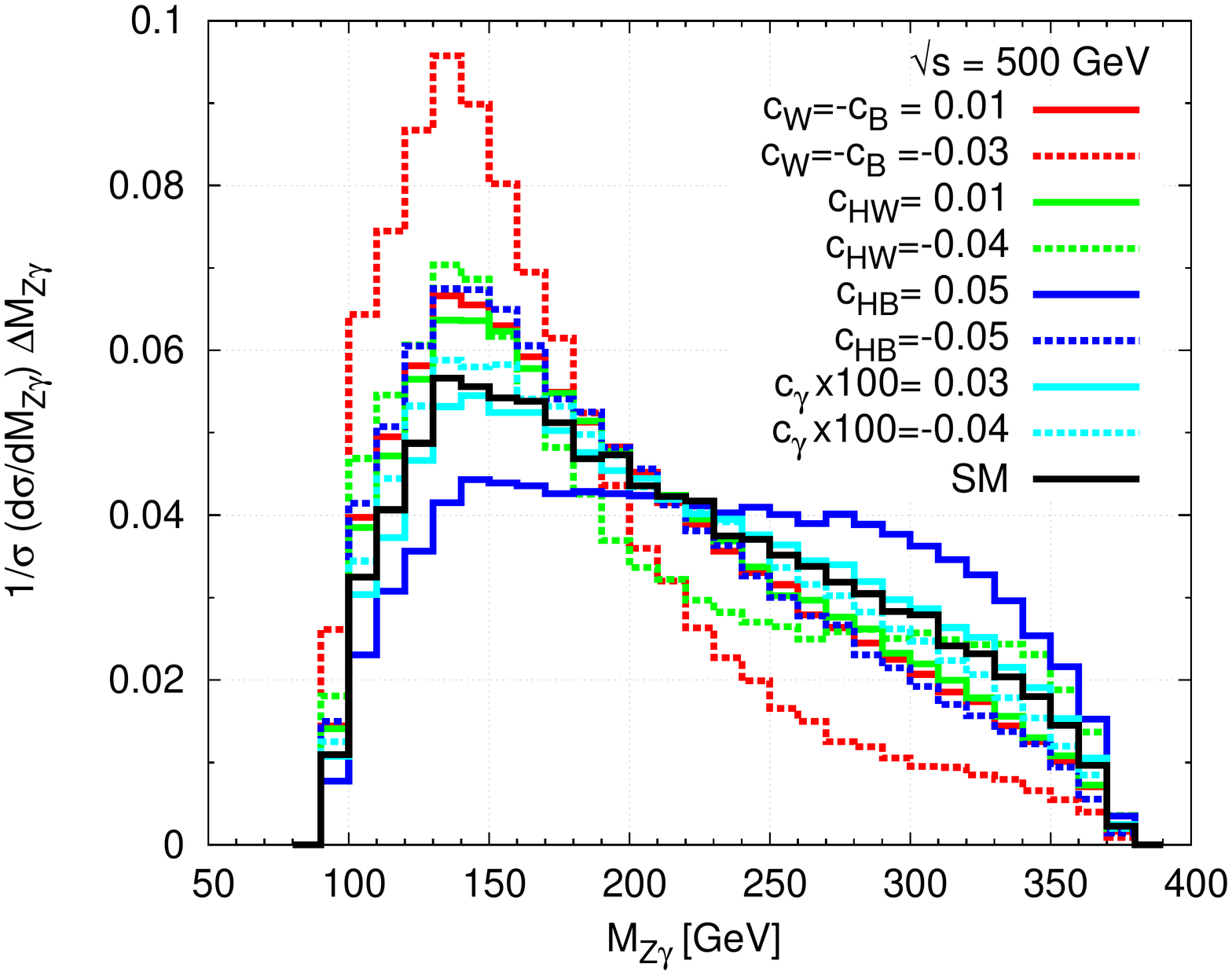} \\
\hspace{1mm}
\includegraphics[angle=0,width=80mm]{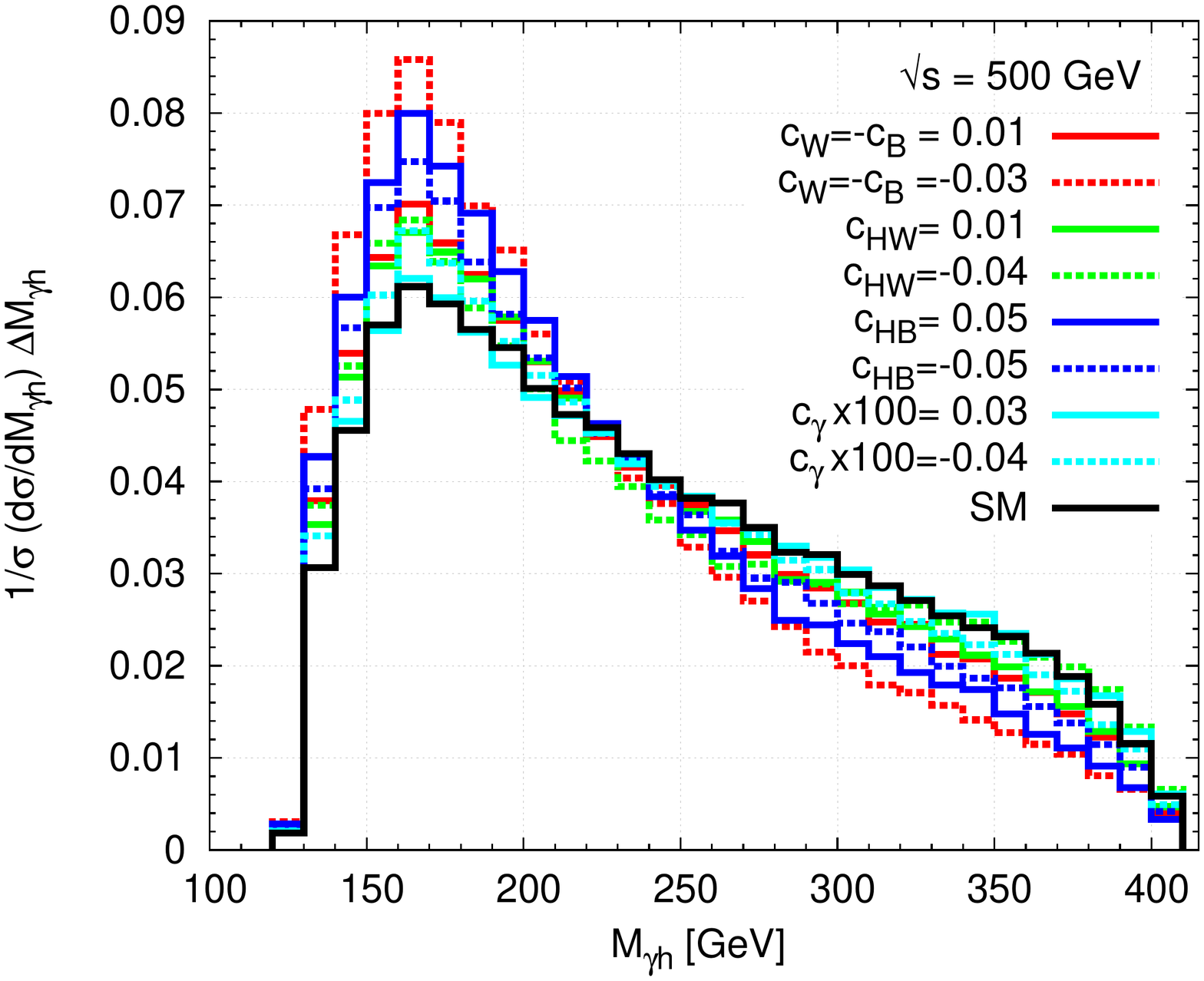} &
\vspace{-8mm}\hspace{-10mm}
\includegraphics[angle=0,width=80mm]{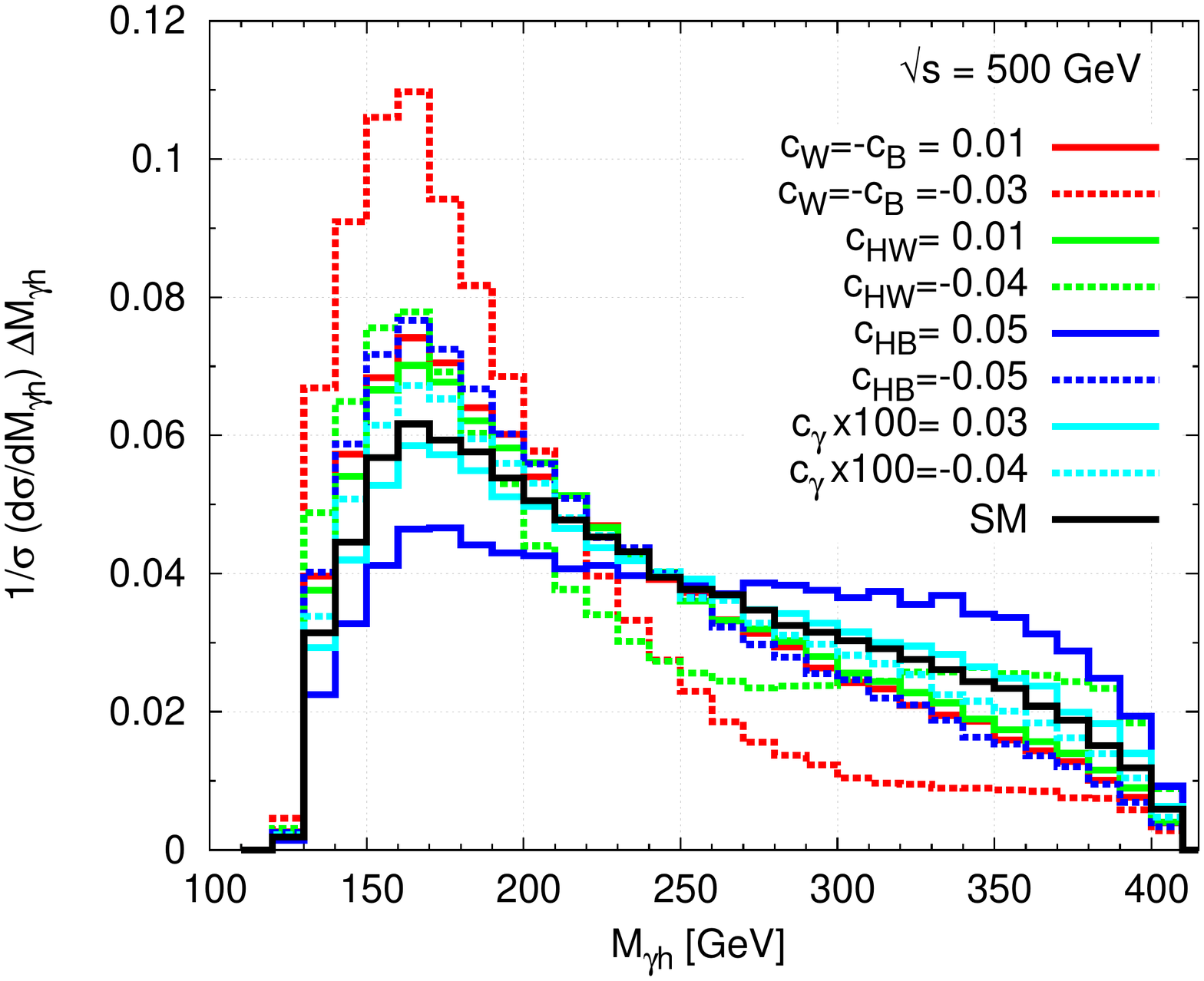} \\
\hspace{1mm}
\includegraphics[angle=0,width=80mm]{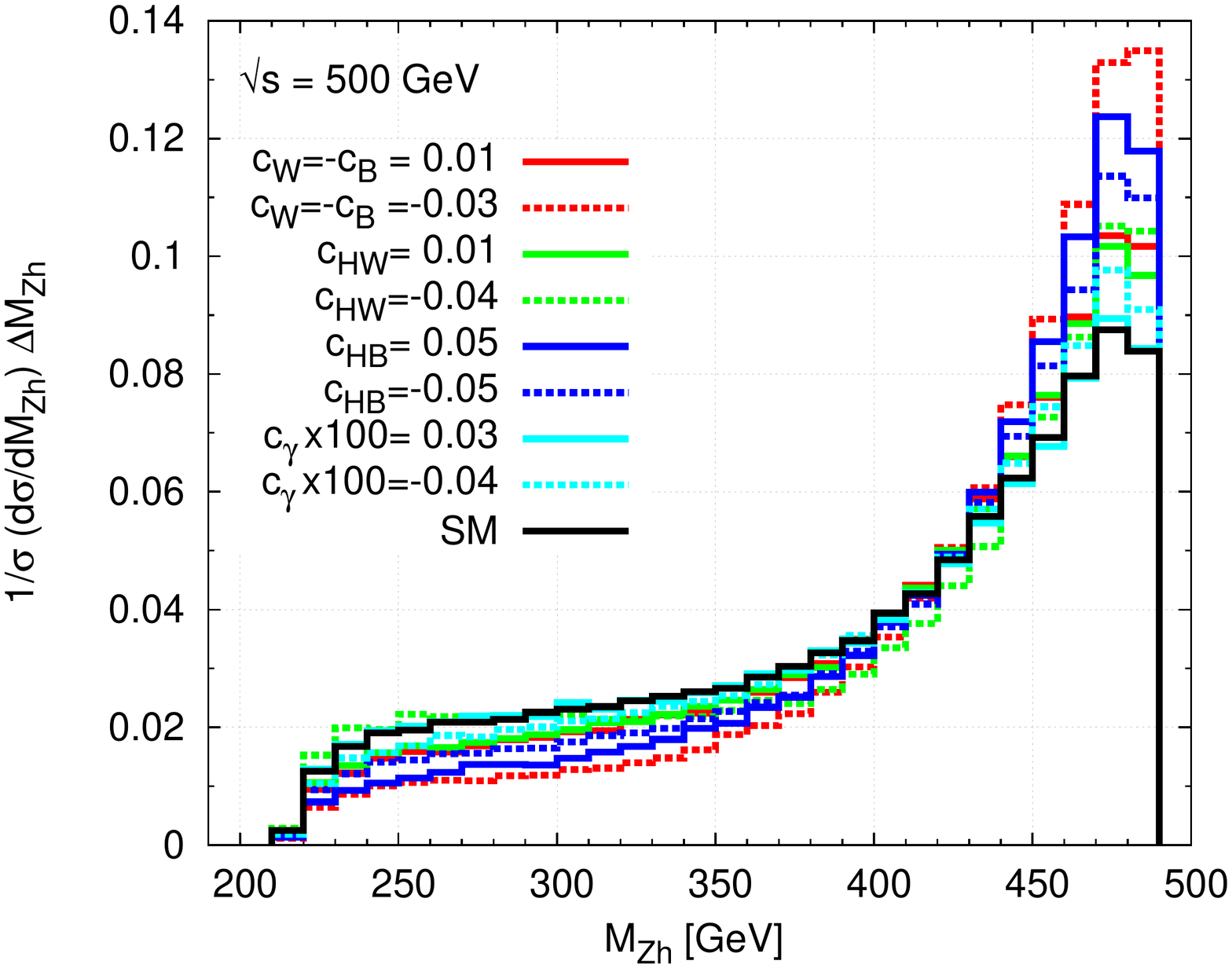} & 
\hspace{-10mm}
\includegraphics[angle=0,width=80mm]{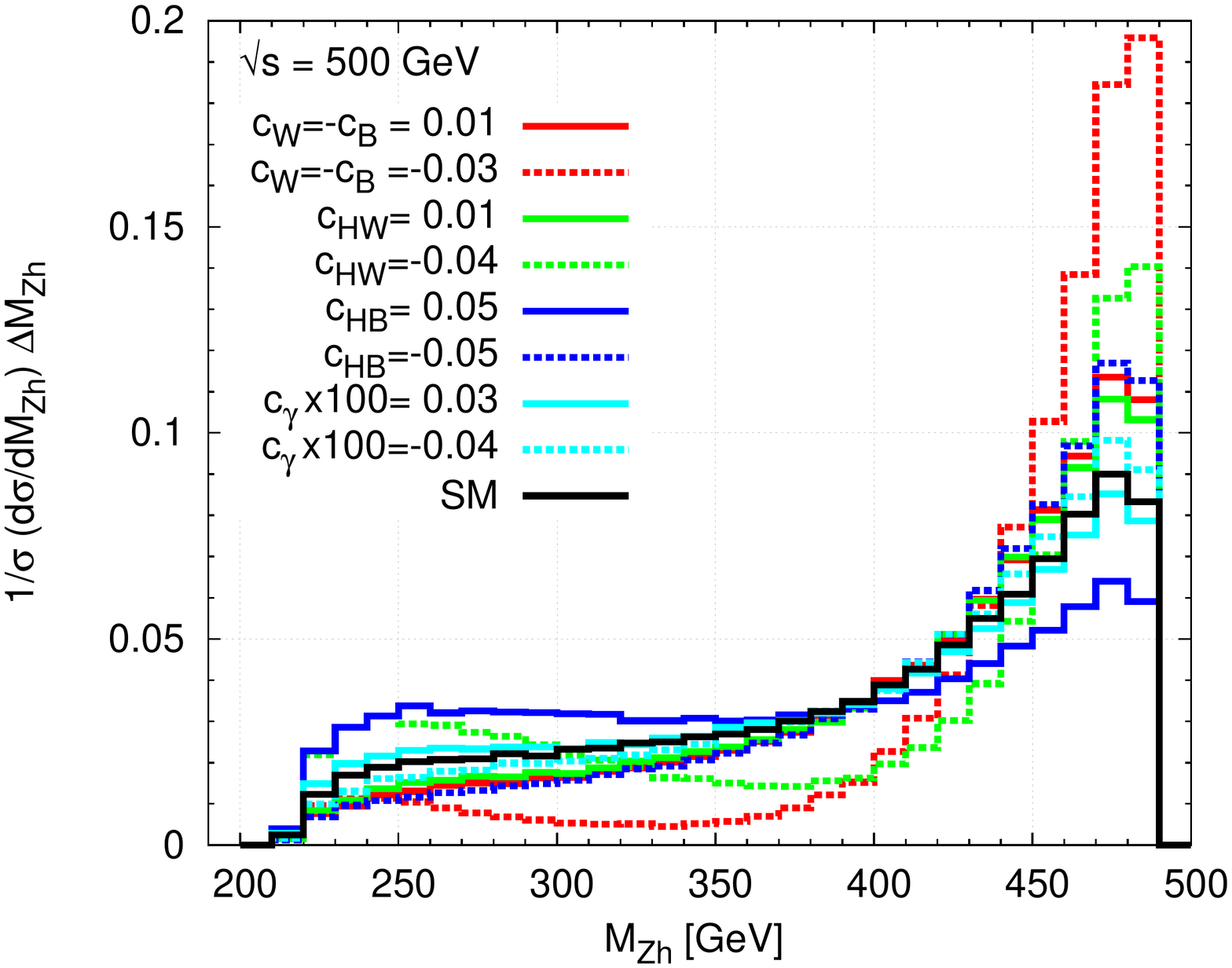} \\
\end{tabular}
\vspace{-10mm}
\caption{The normalized invariant mass distributions of $Z \gamma$ (top), $\gamma h$ (middle) and $Zh$ (bottom) for different anomalous couplings with unpolarized (left) and polarized with $P_{e^-}=-80\%,~~P_{e^+}=+30\%$ (right) beams. A center-of-mass energy of $500$ GeV is assumed.}
\label{fig:invmass}
\end{figure}

The effect of the anomalous couplings on the kinematic distributions are presented in Figs. \ref{fig:angZ}, \ref{fig:angH}, \ref{fig:angZH_AH_ZA}, \ref{fig:Pt}, \ref{fig:energy} \& \ref{fig:invmass}. The $CP$-conserving couplings $\bar c_W$, $\bar c_{HW}$, $\bar c_{HB}$ and $\bar c_{\gamma}$ parameter choices considered for these numerical analyses are \\
\begin{center}
\(
\bar c_W= -0.03, +0.01, ~~~\bar c_{HW}=-0.04, +0.01, ~~~\bar c_{HB}=-0.05, +0.05, ~~~\bar c_{\gamma}=-0.04, +0.03
\)
\end{center}

\noindent
While for $\bar c_W$, the maximum allowed values as per the present bounds are used, in the case $\bar c_{HW}$ or $\bar c_{HB}$ or $\bar c_{\gamma}$, it is somewhat arbitrary but within the limits. While considering beam polarization, an 80\% left-polarized electron beam and a 30\% right-polarized positron beam are assumed, as is expected in the first phase of the ILC, according to the present baseline design. 

We first consider in Fig.\ref{fig:angZ} the normalized $\cos\theta_{Z}$ distributions of the Higgs boson for the SM case, as well as different cases with anomalous couplings as indicated in the figure, while all other parameters are set to zero. The normalized distributions provide clear information on the shape of the distribution, bringing out the qualitative difference between the different cases considered.  The shape of the distribution remains more or less the same as that of the SM case, except a small enhancement in the central regions when $\bar c_{HB} = 0.05$ (solid blue) in the case of unpolarized beams. The advantage of beam polarization is evident (figure on the right) when compared to the corresponding unpolarized (figure on the left) case. Here, the case of negative $\bar c_W$ differs from the other cases. This feature can be exploited to discriminate this case from others.

Figure \ref{fig:angH} (left) presents the normalized $\cos\theta_{h}$ distribution. The postive value of $\bar c_{HB}$ changes the nature of the distribution drastically (solid blue) compared to the SM case (solid black), while all other cases have small deviation. This again can be a useful discriminator of the case. The presence of polarized beams gives a discrimination for the case of negative $\bar c_W$. Figure \ref{fig:angZH_AH_ZA} (top, left) shows the normalized $\cos\theta_{Zh}$ distribution (unpolarized beams), where $\theta_{Zh}$ is the angle between $Z$ and $h$. Here all cases are close to the SM one. While Fig. \ref{fig:angZH_AH_ZA} (top, right) shows the significant change in the shape for the case of negative $\bar c_{HW}$.  
 
Move to the $\cos\theta_{\gamma h}$ (middle) and $\cos\theta_{Z\gamma}$ (bottom) distributions. Fig. \ref{fig:angZH_AH_ZA} shows the normalized angular distributions between the Higgs boson and photon. It can be seen that most of the events clustering in the backward direction for the case of unpolarized (left) as well as polarized beams(right). In Fig. \ref{fig:angZH_AH_ZA}, we present the normalized angular distributions between Z-boson and photon for unpolarized (left) and polarized beams (right). Here, unlike the previous case most of the events accumulated in the forward direction. Both the distributions suggest that the forward-backward asymmetry is a quantitative estimator of the presence of anomalous couplings. The percentage of deviation from the SM case for the cases of a considered set of parameters at fixed center-of-mass energy of 500 GeV with unpolarized and polarized beams is given in Table \ref{table:asymmetryAH500} \& \ref{table:asymmetryZA500}, where the asymmetry is defined as
\begin{equation}
{A_{FB} = \frac{\left[\int_{0}^{1}\frac{d\sigma}{d\cos\theta}d\cos\theta - \int_{-1}^{0}\frac{d\sigma}{d\cos\theta}d\cos\theta\right]}{\left[\int_{0}^{1}\frac{d\sigma}{d\cos\theta}d\cos\theta + \int_{-1}^{0}\frac{d\sigma}{d\cos\theta}d\cos\theta\right]}}
\label{eqn:FBc_asymmetry}
\end{equation}
\begin{equation}
\Delta A_{FB}(\%) = \frac{\left|A_{FB}^{ano} - A_{FB}^{SM}\right|}{A_{FB}^{SM}}\times100.
\label{eqn:percentageasymmetry}
\end{equation}

\begin{table}[H]
\begin{center}
\begin{tabular}{|c|c|c|}
\hline
$Parameter$ &\multicolumn{2}{c|}{\textbf{$\Delta A_{FB}(\cos\theta_{\gamma h}) \%$}}  \\\cline{2-3}
&Unpolarized beams &$P_{e^-}=-80\%$,  $P_{e^+}=30\%$   \\ 
\hline\hline
$c_{HW}$=+0.01      &$19$   &$26$     \\\cline{1-3}
$c_{HW}$=-0.04      &$1$    &$12$     \\\cline{1-3}
\hline
$c_{HB}$=+0.05      &$44$   &$52$     \\\cline{1-3}
\hline
$c_{HB}$=-0.05      &$29$   &$39$     \\\cline{1-3}
\hline
$c_{\gamma}$=+0.03  &$1$    &$8$     \\\cline{1-3}
\hline
$c_{\gamma}$=-0.04  &$13$   &$12$     \\\cline{1-3}
\hline
$c_W=-c_B$=+0.01    &$20$   &$33$     \\\cline{1-3}
\hline
$c_W=-c_B$=-0.03    &$53$   &$77$     \\\cline{1-3}
\hline
\hline
\multicolumn{1}{|c|}{SM~case;  $A_{FB}=$}    &$0.3584$    &$0.3593$    \\\cline{1-3}
\hline
\end{tabular}
\caption{ {Observed forward-backward asymmetry and its deviation from the SM in the  angular distribution ($\theta_{\gamma h}$) at center-of-mass energy of $500$ GeV.}}
\label{table:asymmetryAH500}
\end{center}
\end{table}

\begin{table}[H]
\begin{center}
\begin{tabular}{|c|c|c|}
\hline
$Parameter$ &\multicolumn{2}{c|}{\textbf{$\Delta A_{FB}(\cos\theta_{Z \gamma}) \%$}}  \\\cline{2-3}
&Unpolarized beams &$P_{e^-}=-80\%$,  $P_{e^+}=30\%$   \\ 
\hline\hline
$c_{HW}$=+0.01      &$17$   &$26$     \\\cline{1-3}
$c_{HW}$=-0.04      &$1$    &$14$     \\\cline{1-3}
\hline
$c_{HB}$=+0.05      &$42$   &$52$     \\\cline{1-3}
\hline
$c_{HB}$=-0.05      &$27$   &$38$     \\\cline{1-3}
\hline
$c_{\gamma}$=+0.03  &$3$    &$10$     \\\cline{1-3}
\hline
$c_{\gamma}$=-0.04  &$11$   &$13$     \\\cline{1-3}
\hline
$c_W=-c_B$=+0.01    &$21$   &$33$     \\\cline{1-3}
\hline
$c_W=-c_B$=-0.03    &$51$   &$76$     \\\cline{1-3}
\hline
\hline
\multicolumn{1}{|c|}{SM~case;  $A_{FB}=$}    &$0.3980$    &$0.3954$    \\\cline{1-3}
\hline
\end{tabular}
\caption{ {Observed forward-backward asymmetry and its deviation from the SM in the  angular distribution ($\theta_{Z \gamma}$) at center-of-mass energy of $500$ GeV.}}
\label{table:asymmetryZA500}
\end{center}
\end{table} 

Coming to the other kinematic distributions like the noramlized transverse momentum and energy distributions. In Fig. \ref{fig:Pt}, we present the normalized transverse momentum distributions of the Higgs boson (top), Z-boson (middle) and photon (bottom) with unpolarized (left) and polarized beams (right) for the SM case as well as different cases with anomalous couplings as indicated in the figure, while all other parameters are set to zero. Here again, we find possibilities to distinguish different scenarios with the help of these distributions. It is quite obvious from these distributions that the case of positive $\bar c_{HB}$ or negative $\bar c_W$ appear with quatitative changes, when unpolarized beams considered. While, the qualitative changes can be seen in the case of polarized beams, where the case of positive $\bar c_{HB}$ and negative $\bar c_W$ quite distiguishable from the SM. Next in Fig. \ref{fig:energy}, similar behaviour can be seen in the normalized energy distributions of the Higgs boson (top), Z-boson (middle) and photon (bottom) for the SM along with anomalous couplings as indicated in the figure, with unpolarized (left) and polarized beams (right).    
 
Finally, we consider the normalized invariant mass distributions of $Z\gamma$, $\gamma h$ and $Zh$. Figure \ref{fig:invmass} presents the sensitivity of invariant mass distribution to the anomalous coupling parameters along with the SM at the center-of-mass energy of 500 GeV. This can thus, provide an additional tool to distinguish similar scenarios as in the previous case. 

We notice that in all cases, the beam polarization is found to be useful in terms of improved sensitivity with more than double the number of events compared to the case of the unpolarized beam, while keeping the qualitative features (shape of the curve). Thus, the reach of the probe of the couplings can be improved by a factor of 1.5 to 2 in all cases. Table \ref{table:angle-distinguish} summarizes the cases that could be distinguished from considered kinematics distributions.
 
\begin{table}[H]
\begin{center}
\begin{tabular}{|c|c|c|c|c|c|c|c|c|c|c|c|}
\hline 
&&&&&&&&&&&\\[-4mm]
Couplings&$\cos\theta_h$ &$\cos\theta_Z$ &$\cos\theta_{\gamma h}$ &$\cos\theta_{Zh}$ &$\cos\theta_{Z \gamma}$ &$P_T(h)$  &$P_T(Z)$ &$E(\gamma)$ &$M_{\gamma h}$ &$M_{Z \gamma }$ &$M_{Zh}$ \\[1.5mm] \cline{1-12} 
&&&&&&&&&&&\\[-3mm]
$\bar c_{HW}$  negative alone    &$\times$ &$\times$  &$\times$  &$\checkmark$  &$\times$ &$\checkmark$ &$\checkmark$ &$\checkmark$ &$\checkmark$ &$\checkmark$ &$\checkmark$\\
$\bar c_{HB}$  positive alone    &$\checkmark$ &$\checkmark$ &$\checkmark$ &$\times$  &$\checkmark$ &$\checkmark$ &$\checkmark$ &$\checkmark$ &$\checkmark$ &$\checkmark$ &$\checkmark$\\
$\bar c_{HB}$  negative alone    &$\times$ &$\times$ &$\times$  &$\times$  &$\times$  &$\times$  &$\checkmark$  &$\times$  &$\times$  &$\times$  &$\times$\\
$\bar c_{W}$ positive alone      &$\times$ &$\times$  &$\times$  &$\times$ &$\times$ &$\times$ &$\times$ &$\times$ &$\times$ &$\times$ &$\times$\\
$\bar c_{W}$ negative alone      &$\checkmark$ &$\checkmark$ &$\checkmark$ &$\times$  &$\checkmark$ &$\checkmark$ &$\checkmark$ &$\checkmark$ &$\checkmark$ &$\checkmark$ &$\checkmark$\\[-3mm]
&&&&&&&&&&&\\
\hline
\end{tabular}
\caption{Presence ($\checkmark$) or absence ($\times$) of deviations that could be expected in case of different scenarios with combinations of $\bar c_W$ and $\tilde c_{HW}$ realized from Figs.~\ref{fig:angZ},~\ref{fig:angH},~\ref{fig:angZH_AH_ZA},~\ref{fig:Pt},~\ref{fig:energy} and \ref{fig:invmass}.}
\label{table:angle-distinguish}
\end{center}
\end{table}

\section{Summary and Conclusions} \label{sec:summary}
The discovery of the Higgs boson by the ATLAS and CMS collaborations at the LHC has confirmed the Higgs mechanism as the way to have EWSB providing masses to the fundamental particles. The properties of the Higgs boson measured by the LHC so far are consistent with the expectations of the SM. It is expected that the LHC would measure the mass, spin, and parity of this particle along with the standard decay widths somewhat precisely. On the other hand, details of the couplings like the trilinear and quartic self-couplings as well as the couplings with the gauge bosons are not expected to be measured precisely. At the same time, precise knowledge of these couplings is very important in reconstructing the EWSB mechanism. A precision machine like the ILC is expected to help in the precise measurement of these couplings. In this paper, the process $e^-e^+ \rightarrow Z \gamma h$, which is influenced by the Higgs to gauge boson couplings, namely, $ZZh,~Z\gamma h,~h\gamma\gamma$ and $ZZh\gamma$ is considered. The reach of an ILC at $\sqrt{s}=500$ GeV with an integrated luminosity of 300~fb$^{-1}$ in probing the different relevant parameters of the corresponding effective Lagrangian is presented. The influence of the presence of other couplings in the probe of each of the couplings are studied. We observed that the study of the $\bar c_{HW}-\bar c_{HB}$, $\bar c_{W}-\bar c_{HB}$ and $\bar c_{\gamma}-\bar c_{HB}$ planes show that the allowed region can be narrowed to a very small band. While this band is unaffected by the presence of $\bar c_W>0$, the effect is significant if $\bar c_W<0$.  Consideration of the angular distributions of the Higgs boson ($\cos\theta_{h}$), the Z-boson ($\cos\theta_{Z}$), $\cos\theta_{Z\gamma}$, $\cos\theta_{\gamma h}$ and $\cos\theta_{Zh}$ and the transverse momentum as well as energy distributions of the Higgs boson, Z-boson and photon {have proved} to provide a handle in distinguishing the presence of different scenarios of $\bar c_W$, $\bar c_{HW}$ and $\bar c_{HB}$. All other parameters have an indistinguishable effect on these distributions. The invariant mass distributions of the $Z\gamma$, $\gamma h$ and $Zh$ pairs are also sensitive to some parameters related to anomalous couplings. A quantitative estimate of the forward-backward asymmetry corresponding to the angle between $\gamma$ and $h$ as well as Z and$\gamma$ show that large deviations of up to 77\% are possible for moderate values of the couplings. In all cases, a suitably chosen beam polarization is found to be advantageous, as illustrated with an 80\% left-polarized electron beam and a 30\% right-polarized positron beam. 
The statistics can be improved by a factor of 1.5 with the baseline polarization quoted above, which can be improved to an enhancement factor of 1.7 with the expected 60\% positron beam polarization in the upgraded version of the ILC. Apart from the overall normalizing factor, some details are also affected, as is illustrated in the improvements in the forward-backward asymmetry, when beam polarization is used. Thus, the study shows that $Z\gamma h$ production at the ILC is useful in detecting the anomalous couplings in Higgs-gauge boson interactions. A detailed analysis involving standard kinematic distributions could be used to distinguish different scenarios involving the couplings. While the numerical study needs to be improved with more realistic collider and detector information, as well as study of the background processes, we hope to have conveyed the importance of the process in determining and disentangling the effects of anomalous Higgs-gauge boson couplings.\\

\vskip 5mm
\noindent
{\bf Acknowledgements} SK would like to acknowledge the financial support from the SERB-DST, India, under the National Post-doctoral Fellowship programme, Grant No. PDF/2015/000167 and the IIT-Gandhinagar during the initial stages of the project. We thank B. Coleppa and P. Poulose for useful discussions. 
\\[2mm]

\section{Appendix}
{\bf A.} The most general formula for the matrix element square for arbitrary polarized $e^{-}e^{+}$ beams is 
given in Refs.\cite{Kleiss:1986ct, Hikasa:1985qi}. if one considers massless electron and positron,
the final formula is reduced to the following form \cite{Melic:2016lxl},

\begin{eqnarray}
\left|T\right|^2 &=& \frac{1}{4}\Big[\big(1-P^{L}_{e^-}\big)\big(1+P^{L}_{e^+}\big)\left|T_{e_{L}^{-}e_{R}^{+}}\right|^2+\big(1+P^{L}_{e^-}\big)\big(1-P^{L}_{e^+}\big)\left|T_{e_{R}^{-}e_{L}^{+}}\right|^2 \\\nonumber 
&& +~P^{T}_{e^-}P^{T}_{e^+}~Re\Big(e^{-i(\alpha_{+}+\alpha_{-})}T_{e_{L}^{-}e_{R}^{+}}T^*_{e_{R}^{-}e_{L}^{+}}+e^{i(\alpha_{+}+\alpha_{-})}T_{e_{R}^{-}e_{L}^{+}}T^*_{e_{L}^{-}e_{R}^{+}}\Big)\Big] 
\end{eqnarray}
where $T_{e_{\lambda_1}^{-}e_{\lambda_2}^{+}}$ is the helicity amplitude for the process under consideration, and $\lambda_1$ and $\lambda_2$ are the helicities of the electron and positron respectively. $P_{e^{\mp}}^{L}$ = Degree of longitudinal polarization  and $P_{e^{\mp}}^{T}$  = Degree of transverse polarization of  electrons and positrons. The $\alpha_{\mp}$ refers to the angle of polarization of the electron and the positron, respectively. In our numerical analysis, $\alpha_{+}~+~\alpha_{-}$ is zero as the azimuthal angle of the initial beams is set to zero in {\sc madgraph}. \\

{\bf B.} Throughout the paper, we calculated 3-$\sigma$ limit from the cross section with an integrated luminosity ($\cal{L}$) of 300 $fb^{-1}$ using following formula.  \\[2mm]
$~~~~~~~~~~~~~~~~~~~~~~~~~~~~~~~~~~~~~~n$-$\sigma$ = $\sigma_{SM} \pm n \sqrt{\frac{\sigma_{SM}}{\cal{L}}}$, where $n = 1, 2, 3...$

\end{document}